%% file: manuscript.tex
\documentclass[pre,twocolumn,superscriptaddress]{revtex4-2} 

\usepackage[dvipsnames]{xcolor}
\usepackage{float}
\usepackage{graphicx}
\usepackage{color}
\usepackage{amssymb}
\usepackage{amsmath}
\usepackage{accents}
\usepackage{hyperref}
\usepackage{lipsum}
\usepackage{subfiles}
\usepackage{physics}
\usepackage{tikz}
\usepackage[normalem]{ulem}
\usepackage{pgfplots}
\usepgfplotslibrary{statistics}
\usepgfplotslibrary{groupplots}
\usetikzlibrary{plotmarks}
\usetikzlibrary{arrows.meta}
\usepackage{pgfplotstable}
\usepackage{verbatim}

\tikzset{>=latex}

\pgfplotsset{compat=newest}

\graphicspath{{img/}}

\begin{document}
\title{%
Modelling chromosome-wide target search}
\date{\today}

\author{Lucas Hedstr\"{o}m}
\email{lucas.hedstrom@umu.se}
\affiliation{Integrated Science Lab, Department of Physics, Ume\r{a} University,
SE-901 87 Ume\r{a}, Sweden}
\author{Ludvig Lizana}
\email{ludvig.lizana@umu.se}
\affiliation{Integrated Science Lab, Department of Physics, Ume\r{a} University,
SE-901 87 Ume\r{a}, Sweden}

\begin{abstract}
	\subfile{sections/abstract.tex}

\end{abstract}

\maketitle

\section{Introduction}
\subfile{sections/introduction.tex}

\section{Methods}
\subfile{sections/method.tex}

\section{Results \& Discussion}
\subfile{sections/results.tex}

\section{Conclusions}
\subfile{sections/conclusions.tex}

\acknowledgements%
We acknowledge financial support from the Swedish Research Council (grant no. 2017-03848). We are also grateful to Moa Lundkvist (Ume{\aa} University) for helping with the figures and data handling. This research was conducted using the High Performance Computing Center North (HPC2N), endorsed by the Swedish National Infrastructure for Computing (SNIC), and partially funded by the Swedish Research Council (grant agreement no. 2018-05973). 

\appendix
\subfile{sections/appendix.tex}

\bibliographystyle{unsrt}
\bibliography{refs}

\end{document}

%% file: sections/abstract.tex
%
%
The most common gene regulation mechanism is when a transcription factor protein binds to a regulatory sequence to increase or decrease RNA transcription.
%
%
However, transcription factors face two main challenges when searching for these sequences. First, they are vanishingly short relative to the genome length. Second, there are many nearly identical sequences scattered across the genome, causing proteins to suspend the search. But as pointed out in a computational study of LacI regulation in  \textit{Escherichia coli}, such almost-targets may lower search times if considering DNA looping. In this paper, we explore if this also occurs over chromosome-wide distances.
%
%
To this end, we developed a cross-scale computational framework that combines established facilitated-diffusion models for basepair-level search and a network model capturing chromosome-wide leaps. To make our model realistic, we used Hi-C data sets as a proxy for 3D proximity between long-ranged DNA segments and binding profiles for more than 100 transcription factors. 
%
%
Using our cross-scale model, we found that median search times to individual targets critically depend on a network metric combining node strength (sum of link weights) and local dissociation rates. Also, by randomizing these rates, we found that some actual 3D target configurations stand out as considerably faster or slower than their random counterparts. This finding hint that chromosomes' 3D structure funnels essential transcription factors to relevant DNA regions.

%% file: sections/introduction.tex
When responding to changes in the immediate environment or coordinating embryo development, cells regulate internal protein production. While there are many regulation layers, the most common approach is when DNA-binding proteins, like transcription factors, attach to target sequences to start or stop DNA transcription. However, because these sequences are so short compared to the DNA's length, finding them represents a needle-in-a-haystack problem. In humans, the targets are hundreds of millions of times shorter than the DNA itself
\footnote{Typical Transcription factor sequences are 10-50 bp long. The human genome has $7\cdot10^9$ bps.}.
Despite this problem, protein production works seamlessly and measured search times are relatively short --- about an order of magnitude faster than random search \cite{hammar2012lac}.

Another problem is that there are many almost-targets \cite{marklund2022sequence}. Scattered across the genome, most such targets reside out-of-interaction range from the actual target gene. At first glance, such dispersion should slow the search since DNA-binding proteins get held up at the wrong places. Indeed, this happens in a simple diffusion model where a searcher gets caught in trap after trap. However, if the searcher may intermittently relocate over large distances, e.g.,  via DNA looping, such traps can instead facilitate the search. This seems to occur for the LacI repressor in \textit{Escherichia coli}. As shown in a modeling study~\cite{bauer2015real}, LacI's main target has two flanking auxiliary binding sites that help lower the average search time if LacI first binds one of them and then loops into the primary target.

In this paper, we built a mathematical model to explore whether long-distant auxiliary binding sites could facilitate target search via chromosome-wide loops. We focus on human DNA, where there are potentially thousands of auxiliary binding sites. While most are far from the primary target, counted in the number of base pairs, some auxiliary sites could be close in 3D due to chromatin folding and help channel proteins to their designated DNA regions. This mechanism is arguably hard to measure experimentally, but previous work hint that folded chromatin directs protein traffic. For example, using simulations, one study argued that diffusing transcription factors tend to enrich close to chromatin loop anchors \cite{cortini2018theoretical}. This finding is consistent with so-called Highly Occupied Targets that enrich in regions engaged in long-distance DNA interactions \cite{kvon2012hot}. Another computational study demonstrated that targets embedded in networks of loops have varying search times depending on where in the loop they reside \cite{brackley2012facilitated}

We extend these ideas to a chromosome-wide setting. To achieve this, we take advantage of empirical data sets from Hi-C experiments, providing a realistic description of the large-scale chromosome organization \cite{rao20143d}, and datasets containing binding profiles of more than 100 transcription factors \cite{fornes2020jaspar, korhonen2009moods}. We also developed a diffusion model mixing protein search on short and large scales. On short scales, we use a biophysical model of a two-state searcher~\cite{cencini2017}, resting on established Facilitated diffusion models~\cite{benichou2011intermittent, lomholt2009facilitated, mirny2009protein, felipe2021dna}. On large scales, we use a network model that includes chromosome-wide re-locations \cite{nyberg2021modeling}. Using simulations, we find that the spatial configurations and binding strengths of the auxiliary binding sites strongly affect first-passage times.

%% file: sections/method.tex
    \begin{figure*}
        \includegraphics[width=\textwidth]{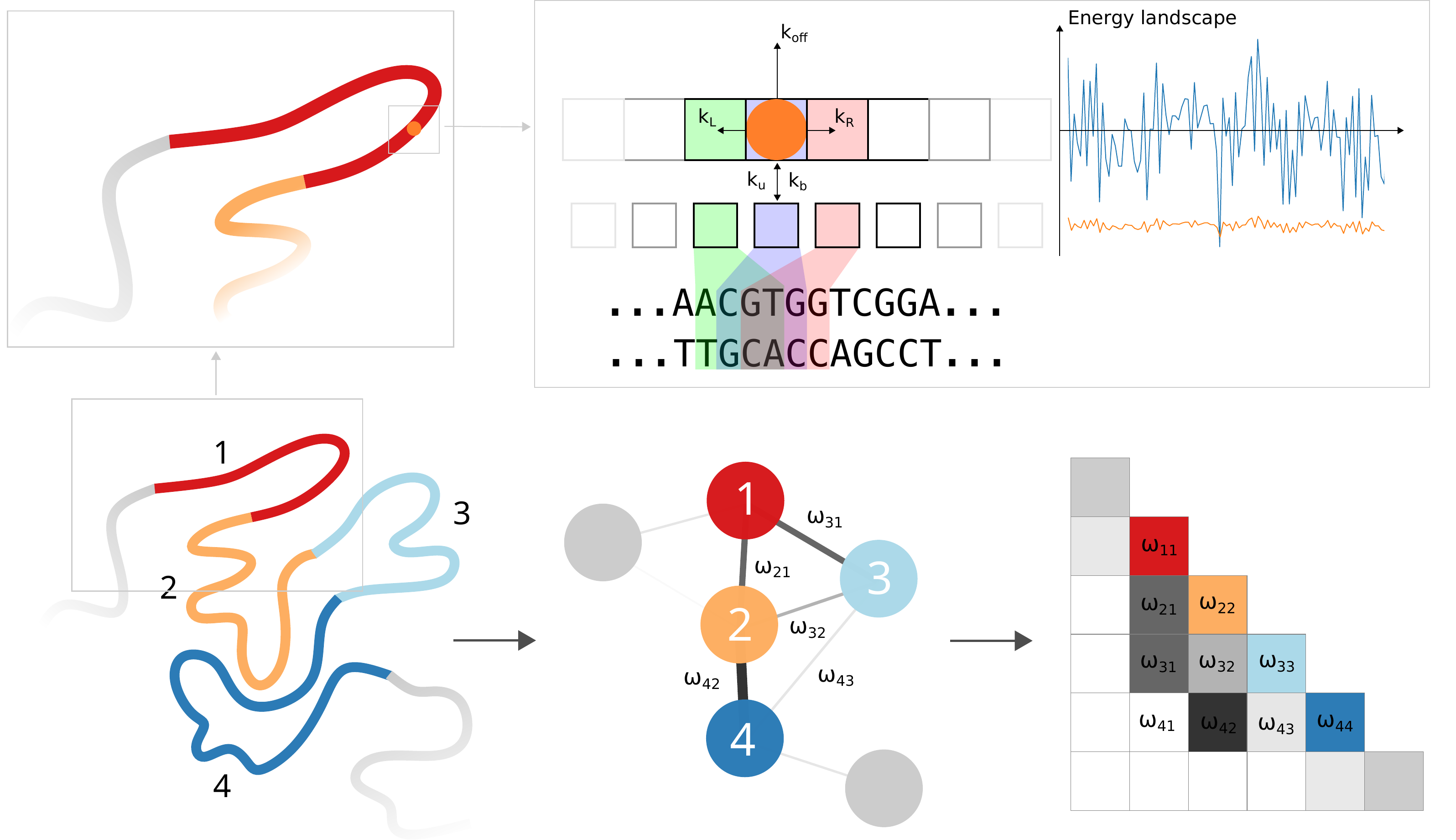}
        \caption{Schematic showing sequence-specific and chromosome-wide search.
        (top) (left) the TF (orange circle) performs sequence-sensitive search along each DNA segment (or Hi-C bin), where it may slide or switch stochastically between two modes --- search and recognition\@.
        (middle) Each box (containing a sequence of length $L^{\rm TF}$) is associated with an energy level reflecting the similarity between the local sequence and TF target motif. We use standard methods (see Sec.\ref{subsec:sequence_sensitive}) to map this similarity onto an energy landscape (right diagram). When in search mode, the transcription factor is less sensitive to the sequence than in recognition mode (blue versus orange lines).
        (bottom) (left) Crumpled DNA polymer. Colours indicate DNA segments representing Hi-C bins.
        (middle) Effective network of 3D interactions between DNA segments. Thin to thick edges represent low to high contact frequencies. (right) we store these frequencies as elements in an adjacency matrix that we use in the simulations. 
        }%
        \label{fig:bp_schematic}
    \end{figure*}
    
    We develop a cross-scale model for site-specific DNA search where we represent a searching transcription factor protein (TF) as a point particle. Our model connects sequence-sensitive search on length scales shorter than a few kilo basepair (kbp) to long mega-basepair (Mbp) leaps across the entire chromosome. To achieve this, we merge established biophysical models for facilitated diffusion, and coarse-grained chromosome-wide search~\cite{cencini2017,nyberg2021modeling}.

\subsection{Sequence-sensitive search}\label{subsec:sequence_sensitive}

    To model search over kbp distances, we use a one-dimensional sliding model designed for TFs looking for 10--20 bps long sequence motifs~\cite{cencini2017}. In this model, the TF switches stochastically between two modes, denoted search and recognition. When in search mode, the TF diffuses along the DNA, while weakly sensing the sequence. But sometimes, the TF occasionally enters recognition mode and stops moving. The more the local sequence resembles the target motif, the higher the probability of switching to recognition mode and the longer it stays bound.
    
    We express the critical rates---transport, binding, and unbinding---in terms of a one-dimensional energy landscape $E^{\rm TF}(x)$ ($x=0,1,2,\ldots$). This landscape represents how well the local sequence $x$ matches the consensus motif; See Fig~\ref{fig:bp_schematic} for an example, and App.~\ref{app:energylandscape} for how we calculate $E^{\rm TF}(x)$. Below follows a brief account of all model's rates.
    
    \textit{1. Diffusion}. We assume that left and right diffusion rates, $k_L(x)$ and $k_R(x)$, depend on the energy difference between neighbouring sites, $x$ and $x\pm 1$:
    \begin{eqnarray}
        k_R(x) &=& k_D e^{\rho(E^{\rm TF}(x) - E^{\rm TF}(x+1))\frac{1}{2}}, 
        \nonumber \\
        k_L(x) &=& k_D e^{\rho(E^{\rm TF}(x) - E^{\rm TF}(x-1))\frac{1}{2}}, 
    \end{eqnarray}
    where $k_B T = 1$, $k_D$ is the natural diffusion frequency, and $\rho$ is a ruggedness factor that reduces the TF's sequence sensitivity when in search mode. Like in~\cite{cencini2017}, we set $\rho=0.2$ and $k_D = 10^7$ to ensure that the average diffusion constant over a long DNA stretch is realistic, \textit{i.e.}, $D=10^7 \, \mathrm{bp}^2 s^{-1} \approx 0.1 \,\mu\mathrm{m}^2 s^{-1}$~\cite{bagchi2008diffusion}. The factor $1/2$ in the exponent guarantees detailed balance: $k_R(x)/k_L(x+1) = e^{-\rho(E^{\rm TF}(x+1)-E^{\rm TF}(x))}$.
    
    \textit{2. Binding and unbinding.} Next, we consider switching between search and recognition mode using two rates: $k_b(x)$ and $k_u(x)$. $k_b(x)$ is the rate to enter recognition mode (binding rate) and $k_u(x)$ denotes the rate to return to the search mode (unbinding rate). Following~\cite{cencini2017}, we assume that these rates depend on the energy differences and obey detailed balance
    \begin{eqnarray}\label{eq:binding_unbinding_rates}
        k_b(x) &=& \gamma e^{(\rho E^{\rm TF}(x) - E^{\rm TF}(x))\frac{1}{2} - \Delta G}, 
        \nonumber \\
        k_u(x) &=& \gamma e^{(E^{\rm TF}(x) - \rho E^{\rm TF}(x))\frac{1}{2}}.
    \end{eqnarray}
    Here, $\gamma = 10^7 s^{-1}$ (representing a fast base transition rate~\cite{murugan2010theory, mirny2009protein}), and $\Delta G$ is the free energy change associated with switching between search and recognition mode. As in \cite{cencini2017}, we calibrate this parameter by letting the energy difference between bound and unbound state be zero for the weakest consensus target along the whole chromosome, $\Delta G = -(1 - \rho_{\rm max})\max E^{\rm TF}(x\in \text{consensus\ targets})$ (see Sec. \ref{subsubsec:moods} for how we define consensus targets). $\rho_{\rm max} = 0.3$ is the largest considered $\rho$ to allow for effective diffusion.
    \footnote{If increasing $\rho$ beyond 0.3, the diffusion constant becomes too low than observed in experiments \cite{cencini2017})}.
    
    Finally, the TF may dissociate from the DNA when in search mode. We define the off-rate as
    \begin{equation}\label{eq:off_rate}
        k_{\mathrm{off}} = \delta e^{\rho E^{\rm TF}(x)},
    \end{equation}
    where $\delta = 10^3 s^{-1}$. In our simulations, this parameter choice corresponds to the sliding length $\approx 90$ bp, which agrees with studies (refs. \cite{hammar2012lac, winter1981diffusion} suggests that it is $\lesssim 100$~bp).
    
    \textit{3. Inter-segmental transfer.} In the cross-scale model, we include loop-assisted jumps between distant DNA sites. And similar to other studies~\cite{bauer2015real, felipe2021dna}, we assume that this process depends critically on the free energy cost of forming a loop. However, we differ from these works in how we calculate this cost. If assuming that DNA forms loops like a polymer, say a Gaussian or worm-like chain, the loop-size distribution decays as a power law and the free energy becomes $\Delta G_l = -\alpha \ln l$, where $l$ is the loop size and $\alpha$ is the looping exponent ($\alpha = 3/2$ for a Gaussian polymer). However, as we know from Hi-C data analysis \cite{dixon2012topological, lee2019mapping}, the contact probabilities over chromosome-wide scales do not decay as simple power-laws \footnote{We acknowledge that the genome-averaged contact probability in human decays as $l^{-1.08}$ \cite{lieberman2009comprehensive} or $l^{-0.75}$ within TADs \cite{sanborn2015chromatin}. However, individual pairwise contacts show significant deviations from these average relationships.}. 
    Instead, Hi-C experiments show that the pairwise contact counts are more complex. Therefore, we treat DNA looping separately in the chromosome-wide search where we take advantage of Hi-C data sets. In short, whenever there is an unbinding event (with the rate $k_{\mathrm{off}}$) the particle translocates to another site on the chromosome with a probability that is proportional to the number of measured chromosome contacts. This includes rebinding to the same segment, mimicking typical intersegmental transfer but to a random location. We refer to Sec.~\ref{subsec:chromosome_wide_search} for details.
    
\subsection{Defining consensus targets and auxiliary binding sites}\label{subsec:energy_profile_and_targets}

    In this paper, there is a critical difference between consensus and auxiliary binding sites. While consensus targets are experimentally-derived TF binding positions, auxiliary binding sites come from motif-matching algorithms that scan the genome sequence for potential binding sites. Interestingly, these two data sets do not always overlap, highlighting that some binding sites lack sequence similarity with the consensus motif \cite{farnham2009insights}.
    
    \subsubsection{Consensus targets} 
    We downloaded the consensus site positions for each TF from the JASPAR database~\cite{jaspar, fornes2020jaspar}. JASPAR is a library of manually curated TF binding profiles from experiments (stored as position frequency matrix). We annotate here the unique TFs by their respective PFM matrix number~\cite{jaspardocs}.
    
    \subsubsection{Auxiliary binding sites}\label{subsubsec:moods}
    
        To computationally determine the auxiliary binding sites, we use the Motif Occurrence Detection Suite (MOODS)~\cite{korhonen2009moods} that calculates the binding score $W^{\rm TF}(x)$ at any genomic position $x$. The binding score is a sum over the position weight matrix (PWM)~\cite{stormo1982use} across positions in a sliding window $L^{\rm TF}$ with the sane length as the potential target
        \begin{equation}
            W^{\rm TF}(x)
            = \sum_{i=x}^{x+L^{\rm TF}} \log{\frac{P^{\rm TF}(a_i, i)}{\pi(a_i)}}.
        \end{equation}
        where $P^{\rm TF}(a_i, i)$ is the position weight matrix [Eq~\eqref{eq:pfm}], and $a_i$ denotes the basepair at position $i$ (A, C, T or G). Also, MOODS separates significant from insignificant auxiliary binding sites using a threshold $T$, where $W^{\rm TF}(x)>T$. This threshold is associated with the probability $p_m$ that a sequence snippet is randomly generated from a background basepair frequency $\pi(a)$.  We used $p_m = 10^{-4}$ and $\pi(a_i)=1/4$  (default settings).
    
\subsection{Chromosome-wide search}\label{subsec:chromosome_wide_search}

    In the sequence-sensitive search, the protein diffuses along DNA until finding a consensus site or dissociating. This section describes how we model rebinding to a potentially distant DNA segment after dissociation.
    
    Rebinding depends critically on the spatial arrangement of DNA segments. While not a part of the classical facilitated diffusion model, several theoretical and experimental studies showed that search times change with correlated rebinding and the type of DNA conformations \cite{slutsky2004kinetics, van2008dna, nyberg2021modeling, amitai2018chromatin, bauer2015real, hu2006proteins, lomholt2009facilitated, lomholt2005optimal, smrek2015facilitated}. Here, we build one of these approaches that circumvent knowing the chromatin's explicit large-scale 3D structure~\cite{nyberg2021modeling}. Instead, it treats DNA's 3D interactions as a contact network and infers contact probabilities between DNA segments from experimental Chromosome Conformation Capture data (we use Hi-C data from~\cite{rao20143d} and LiftOver to UCSC hg38 from UCSC hg19 \cite{kent2002human}). This network approach embraces the full complexity of chromosomes' semi-hierarchical organization that cannot be described using Gaussian or Fractal globule polymers with simple power-law decaying loops-sizes. 

    
    \subsubsection{Network model}
        
        We consider chromatin's 3D interactions as a weighted network where the nodes and links represent DNA segments and 3D contacts, respectively. The links' weights are proportional to the segment-segment contact probabilities derived from Hi-C data. To model the chromosome-wide protein search, we consider a point particle jumping randomly across the network (see Fig~\ref{fig:bp_schematic}). These jumps are associated with rates $\omega_{ij}$ ($i,j = 1,\dots, N$) where $N$ is the number of nodes (typically $N\sim 10^4-10^5$). We calculate $\omega_{ij}$ as the probability $q_{ij}$ to jump between segments $i$ and $j$ multiplied by the frequency of a successful jump $f_\mathrm{coll}$ (`collision frequency'). As a proxy for $q_{ij}$, we use the number of Hi-C contacts $v_{ij}$ (KR-normalized), leading to \cite{nyberg2021modeling}
        \begin{equation}\label{eq:hicrates}
            \omega_{ij} = f_\mathrm{coll} v_{ij}, \ \  \omega_{ij} = \omega_{ji}.
        \end{equation}

        For clarity, we use population-averaged Hi-C data to calculate $\omega_{ij}$. Therefore, the rates do not include cell-to-cell variability and potentially transient loops. Instead, the network model treats the chromosomes as a rigid structure where the probability to jump from one segment to another is proportional to the contact frequency.
        
     \subsubsection{Hi-C data treatment}
    
        To extract the pair-wise DNA contacts, we used Hi-C data from human cell line GM12878~\cite{rao20143d}, downloaded from the GEO database~\cite{edgar2002gene}. Before calculating the rates $\omega_{ij}$, we normalized the contact counts $v_{ij}$ using the Knight-Ruiz algorithm (KR~\cite{knight2013fast}) implemented in \texttt{gcMapExplorer}~\cite{kumar2017genome}. To omit low-frequency contacts and reduce noise, we use the pre-processed data set \texttt{MAPQGE30}. We use data from Chromosome 19, which is the most transcriptionally active chromosome
        \footnote{Close to 80\% of the genes are active, which is significantly more that the other chromosomes where about 50\% are active \cite{nyberg2021modeling}.}.
        
        In the chromosome-wide search, we did not consider chromosome-chromosome contacts. Admittedly, this is an approximation as there are interactions with other chromosomes, albeit much less frequent. To check this assumption, we calculated the ratio of internal versus external contacts from Hi-C data \cite{nyberg2021modeling}. Depending on the chromosome about 75\%–90\% of the contacts are internal. However, because of the low signal-to-noise ratio \cite{kaufmann2015inter}, this number should be taken with caution.
        
        Furthermore, we used 5 kb Hi-C data. 5kb represents a threshold separating sequence-sensitive and chromosome-wide search in our cross-scale model. Admittedly, this threshold is arbitrary, and there are two main reasons for this choice. First, 5 kb agrees with length scales used in other facilitated-diffusion studies that typically vary between 1--10 kb (e.g.~\cite{cencini2017, bauer2015real}). Second, on length scales much larger than 5 kb, say 50--100 kb, the looping probabilities do no longer follow simple power-law relationships. Instead, on these scales, chromatin interactions are more complex. 
        Balancing these arguments, we picked 5 kb as a conservative choice. But to be clear, this choice do not reflect any inherent limitation in our approach. Our cross-scale framework can handle genome-wide Hi-C matrices at any resolution that may include intra- and inter-chromosomal contacts.

    
    \begin{figure*}
        \includegraphics[width=\textwidth]{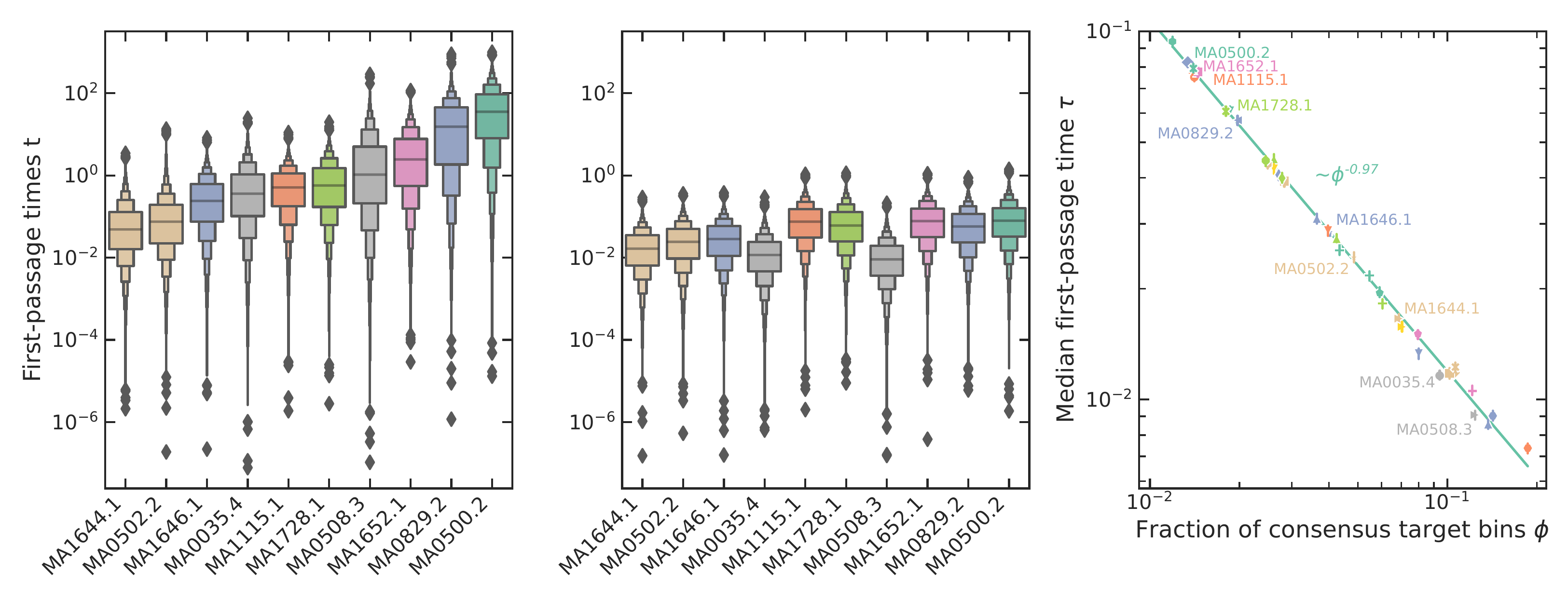}
        \caption{%
        Target bin finding times for 10 transcription factors (TFs) searching for any consensus target bins on chromosome 19 (TF labels: MA1644.1 MA0502.2 etc.). 
        (left) Letter-value plots  \cite{letter-value-plot} showing the distribution of first-passage times to any of the consensus sites. For each TF, we simulated $\sim 10^6$ search times from random starting points. When plotted in log scale, we note that the search times vary by several orders of magnitude (within and  between TFs).
        (middle) Letter-value plots of search times with normalized energy landscapes (see App.~\ref{app:genericsearcher}). Comparing the left and middle panels, we see that the variability between TFs becomes smaller. Yet, there are still significant differences.
        (right) Fraction of target bins containing at least one consensus target versus the median first-passage time (normalized energy landscape). Here we show simulation results from 60 TFs, where marks indicate the TFs in the middle graph. The green line shows the fit $\phi^{-0.97}$ (note the logged axes). This panel indicate that the median search times depends mainly on $\phi$ after we normalize the energy landscape.
        }%
        \label{fig:global_search}
    \end{figure*}
    
\subsection{Cross-scale computational approach}\label{subsec:cross_scale_search}
    
    We combined sequence-sensitive and chromosome-wide search in a cross-scale computational framework to simulate TF search over large distances. This frameworks build on the Gillespie algorithm \cite{gillespie1976general} to calculate transition probabilities and time step lengths. Below follows the main steps

    \begin{enumerate}
    \item 
    We start with a TF in a randomly selected Hi-C bin (node). We exlude isolated bins (where the row sum of the Hi-C matrix is non-zero), such as loci on the centromere and the telomeres. 
    \item 
    Once assigned a Hi-C bin, the TF starts diffusing from a random locus on the corresponding 5 kbp segment. Here it performs sequence-specific search as it diffuses through the energy landscape $E^{\rm TF}(x)$. If encountering deep valleys, the TF may fall into recognition mode and stay trapped for some time ($\propto 1/k_u(x)$, Eq. \eqref{eq:binding_unbinding_rates}). 
    \item
    The TF may also dissociate from DNA (with the rate $\sim k_{\text{off}}$). If so, it will instantaneously relocate to another Hi-C bin, say from bin $i$ to $j$, with the relocation probability  $\omega_{i,j}$ [see Eq. \eqref{eq:hicrates}]. This intermittent transition includes rebinding to the same Hi-C bin, but at a random position $x'$. Modelling rebinding in this way approximately captures facilitated diffusion, albeit with a uniform jumping  probability rather than a power-law
    \footnote{We implemented intermittent power-law jumps within the 5kb region too, but we did not notice any meaningful differences when studying large-distance search times.}.
    \item
    We repeat these steps until the TF reaches a Hi-C bin that contains at least one consensus target site (i.e., consensus bins act as absorbing targets). After storing the search time, we initiate a new run where the TF starts from another random Hi-C bin.
    \end{enumerate}
    
    To speed up the simulations, we made considerable data pre-processing. Foremost, prior to the simulations, we parsed the chromosome's sequence and calculated all TF-specific rates---$k_L$, $k_R$, $k_u$, $k_b$, and $k_{\text{off}}$ [see Eqs. \eqref{eq:binding_unbinding_rates}-\eqref{eq:off_rate}]---and stored them in a separate file that we load once the simulation starts. By calculating the rates in this way, we significantly reduced the computational time. Particularly, this sped up the switch between chromosome-wide and sequence-specific search.

%% file: sections/results.tex
\subsection{Search times to consensus target bins vary among transcription factors} \label{subsec:global_search}
    
    To study search-time variations among TFs, we downloaded 108 TF binding profiles from \cite{jaspar}, annotated by their matrix IDs \cite{jaspardocs}. For each TF, we curated lists of consensus targets and auxiliary binding sites and assigned these to their designated 5 kb Hi-C bins (nodes in the DNA contact network). We noted that most bins contain zero or one consensus target, but some have more than ten. Next, we simulated the search for each TF type separately using our cross-scale model (see Sec. \ref{subsec:cross_scale_search}). Across $> 10^6$ simulation runs with different starting conditions, we recorded the first-passage times to any bin containing at least one consensus target.
    
    Because the genomic locations and the number of consensus targets differ significantly, we expect to observe a wide search-time variability between the TFs. Indeed, plotting the distribution of the $10^6$ search times for 10 TF types (Fig.~\ref{fig:global_search} left panel), we see that some TFs are faster than others, where the medians (horizontal lines) may differ by almost two orders of magnitude. To better understand what causes these variations, we analyzed how search times change with two metrics: effective diffusion constants and the number of consensus target bins. 
    
    As explained in Methods, we envision the TF diffusing through an energy landscape when sliding along DNA. This landscape has valleys and hills reflecting how close the local sequence reassembles the consensus target motif. If there are many local minima, we expect the TF will get stuck repeatedly, thus leading to a lower diffusion constant. For example, the effective diffusion constant associated with a random one-dimensional landscape is approximately $D = D_0 {(1 + \sigma_E^2/2)}^{1/2}e^{-7\sigma_E/4}$, where $\sigma_E^2$ is the landscape's variance and $D_0$ is the diffusion constant when it is flat~\cite{slutsky2004kinetics}. In other words, rougher energy landscapes lead to lower diffusion constants. Therefore, to correct for different $\sigma_E^2$ when comparing TF search times, we renormalized the landscapes' variance to unity (the mean was already zero) and made new simulations (see App.~\ref{app:genericsearcher} for details). We found that the variations are much smaller after re-normalization---compare the left and middle panels in Fig.~\ref{fig:global_search}. However, the median search time still differs by a factor of 10 between the fastest and slowest TF type. 
    
    Apart from varying effective diffusion constants, the number of consensus targets and auxiliary binding sites also change. For example, comparing two TFs having short or long target motifs, say 10 and 20 bp, it is clear that the one looking for the 10 bp motif encounters many more almost-targets in a long random basepair sequence than the one searching for a 20 bp target. Therefore, if tracking search times to any consensus target bin, we expect that TFs having high target counts will have short average search times. 
    
    But there is yet another factor worth considering. While the total target number is important, it is also critical to how far the targets are from each other along the DNA.  For example, if several targets lie within the TF's sliding length, i.e., distance covered before dissociating, it will likely find any of them if in proximity. Therefore, dense target clusters may act as effective targets (see theoretical analysis in App.~\ref{app:effectivetargets}). Furthermore, it also matters how these effective targets distribute in the DNA contact network. For example, if spread out evenly, we expect that the average search time is shorter than if constrained to a few Hi-C bins. To study such effects, we plotted the fraction $\phi$ of Hi-C bins having at least one consensus target against the median-first passage time $\tau$ to any one of these target bins for 40 TFs [Fig.~\ref{fig:global_search} (right)]; we used the normalized energy landscape as in panel (b). We find a convincing inverse relationship, where the search time decays as $\phi^{-0.97}$ (linear fit, solid line).

    To better understand this relationship, we considered a simple search model
    \footnote{A full derivation not assuming a fully-connected network is shown in App~\ref{app:tauonnormalizedhic}.}:
    a fully connected weighted network (even if dense, Hi-C is  not fully connected). If starting at some node $i$, the probability that a searcher finds the target in one step $P_i(1)$ equals the probability that the next node $j$ is a target ($\phi$) times the probability of translocating from $i$ to $j$ ($\omega_{ij}$). Since there are many target nodes in the network, we sum over all $j$,
    \begin{equation}\label{eq:p1}
        P_i(1) =  \sum_j \omega_{ij} \, \phi= \phi,
    \end{equation} 
    where we used that all weights sum to unity $\sum_j \omega_{ij} = 1$ (they are KR-normalized). Generalizing Eq.\eqref{eq:p1} to $n>1$ steps yields
    \begin{equation} \label{eq:P_i}
        P_i(n) = (1-\phi)^{n-1}\phi.
    \end{equation} 
    Since the search trajectories are similar in this toy model, averaging over initial starting points $i$ gives the  geometric distribution $\langle P_i(n) \rangle = P(n) = (1-\phi)^{n-1}\phi$. This distributions has the median $\tilde{n}$~\cite{raade1999mathematics}
    \begin{equation}
        \tilde{n} = \left \lceil \frac{-1}{\log_2(1-\phi)} \right \rceil,
    \end{equation} 
    in which $\log_2$ is the base-2 logarithm and $\lceil \cdot \rceil$ is the round-up function. Since  $\phi < 10^{-1}$ for most TFs, [see Fig. \ref{fig:global_search} (right)], we expand the enumerator for small $\phi$, yielding
    \begin{equation}
        \log_2(1-\phi) = -\frac{\phi}{\log(2)} - \mathcal{O}(\phi^2)
                       \approx -\frac{\phi}{\log(2)}
    \end{equation} 
    and thus
    \begin{equation}
        \tilde n \approx \left \lceil \frac{\log(2)}{\phi} \right \rceil
        \sim \frac{1}{\phi}.
     \end{equation} 
     In the normalized systems, the effect of auxiliary binding sites is not negligible, but considerably weaker compared to the non-normalized systems. For multiple trajectories and long search paths ($n \gg 1$), the search time is proportional to the number of jumps, giving us
    \begin{equation}
        \tilde{t} = \tau \sim \tilde{n} \sim \phi^{-1}
    \end{equation}
    which is consistent with the data in Fig.~\ref{fig:global_search}.

    To conclude, when considering search times to any one of the consensus target bins, we can explain most of the variation with effective diffusion constants and by how much the effective consensus targets are distributed over the chromosome. But even if these averages have moderate variability (after normalization), search times for single target bins may differ substantially from each other. This is the next section's topic.
    
    \begin{figure}
        \includegraphics[width=\columnwidth]{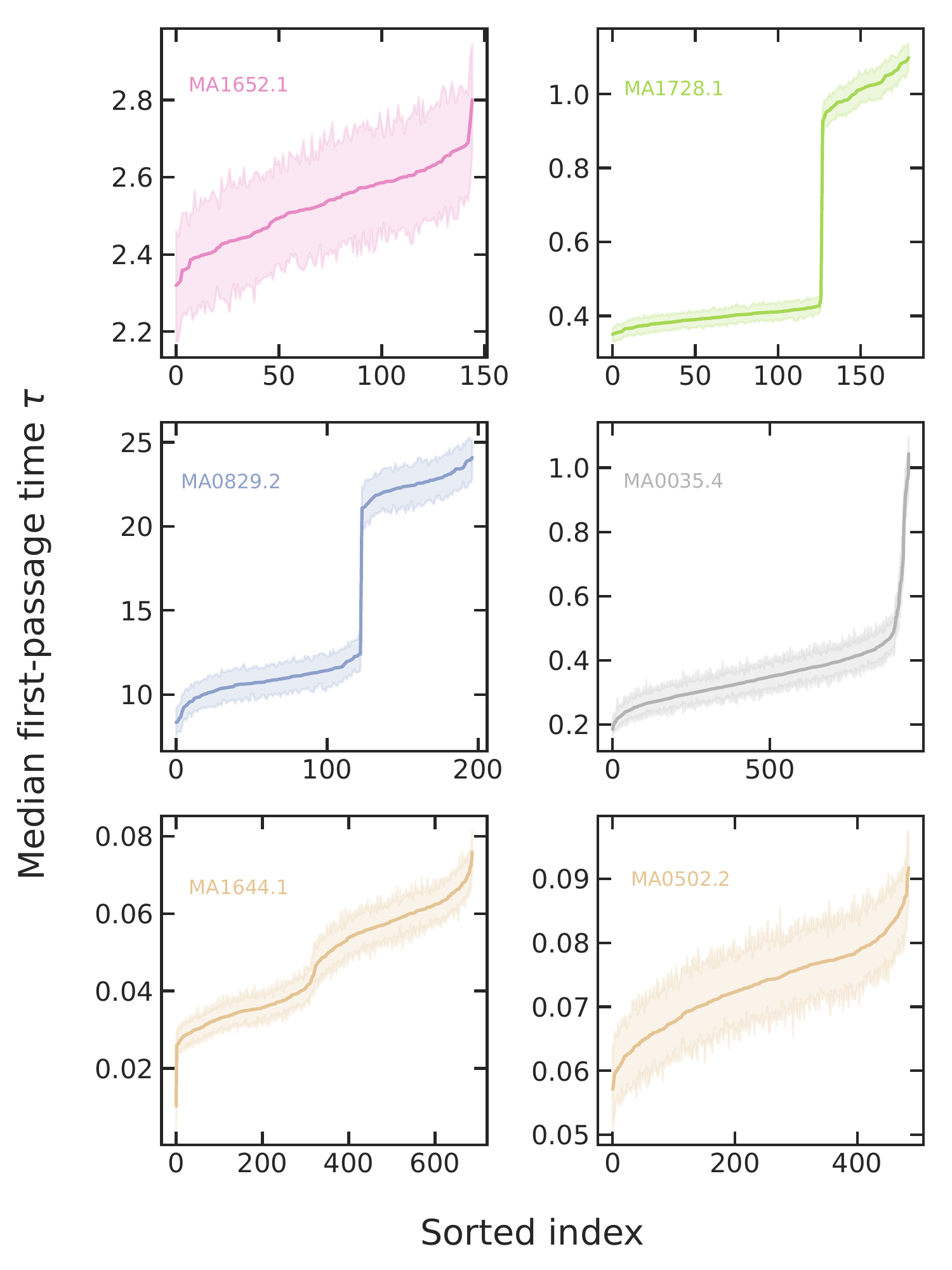} 
        \caption{%
        Sorted median search times $\tau$ to consensus target bins for six TFs. Each panel shows the median first-passage time $\tau$ (solid) with the corresponding 95\% CI (shaded area). Each data point is calculated from $\geq 1000$ realisations from random starting positions on the Hi-C network. While some TFs show larger variation than other, overall we note that $\tau$ varies significantly between consensus sites. Some TFs even has step-like curves (MA0829.2 and MA1728.1) This shape reflects that targets are unevenly distributed between the chromosome arms.
        }%
        \label{fig:tau_sorted_6tfs}
    \end{figure}


\subsection{Search time variability between individual consensus target bins}

    The previous section showed that search times vary if considering all consensus bins as valid targets (Fig.~\ref{fig:global_search}). But in actual gene regulation, some targets are likely more important than others. Some are also easier to find depending on their chromosomal positions. To study search time variability among individual consensus targets, we shift to a microscopic view and analyze the TF's search to specific bins. Just as the Lac repressor gets help from surrounding auxiliary binding sites to quickly find its main operator site, as discussed in the Introduction, we are interested in what extent consensus targets occupy easy-to-access locations on the chromosome and if surrounding auxiliary binding sites arrange in favourable network that help TFs reach specific DNA regions having strong targets or high target densities.

    To quantify this search-time variability, we calculated the median search times $\tau$ to consensus target bins for six TFs (Fig.~\ref{fig:tau_sorted_6tfs}). Each figure panel portrays sorted search times, one TF per panel, where the solid lines represent the median $\tau$ and the shaded areas show the 95\% confidence interval (of $\sim 10^3$ search paths). The panels show that $\tau$ in some cases vary by a factor of 2---5, which is slightly lower than the variation among the TFs themselves (Fig. \ref{fig:global_search}).
    
    We also note that the $\tau$ curve associated with two TFs has an unusual step-like shape relative to the others (MA0829.2 and MA1728.1). This is because the target distribution (both auxiliary and consensus) is skewed between the chromosome arms. If starting on one side, TFs have difficulty to cross to the other because the centromere acts as a barrier, leading to shorter search times on one side than the other. This observation manifests even on the macroscopic TF level, where we observe a broad $\tau$ distribution (e.g.,  MA0829.2 in Fig.~\ref{fig:global_search}).

\subsection{Node flux explains the median first-passage time to most target bins}


    \begin{figure}
        \includegraphics[width=\columnwidth]{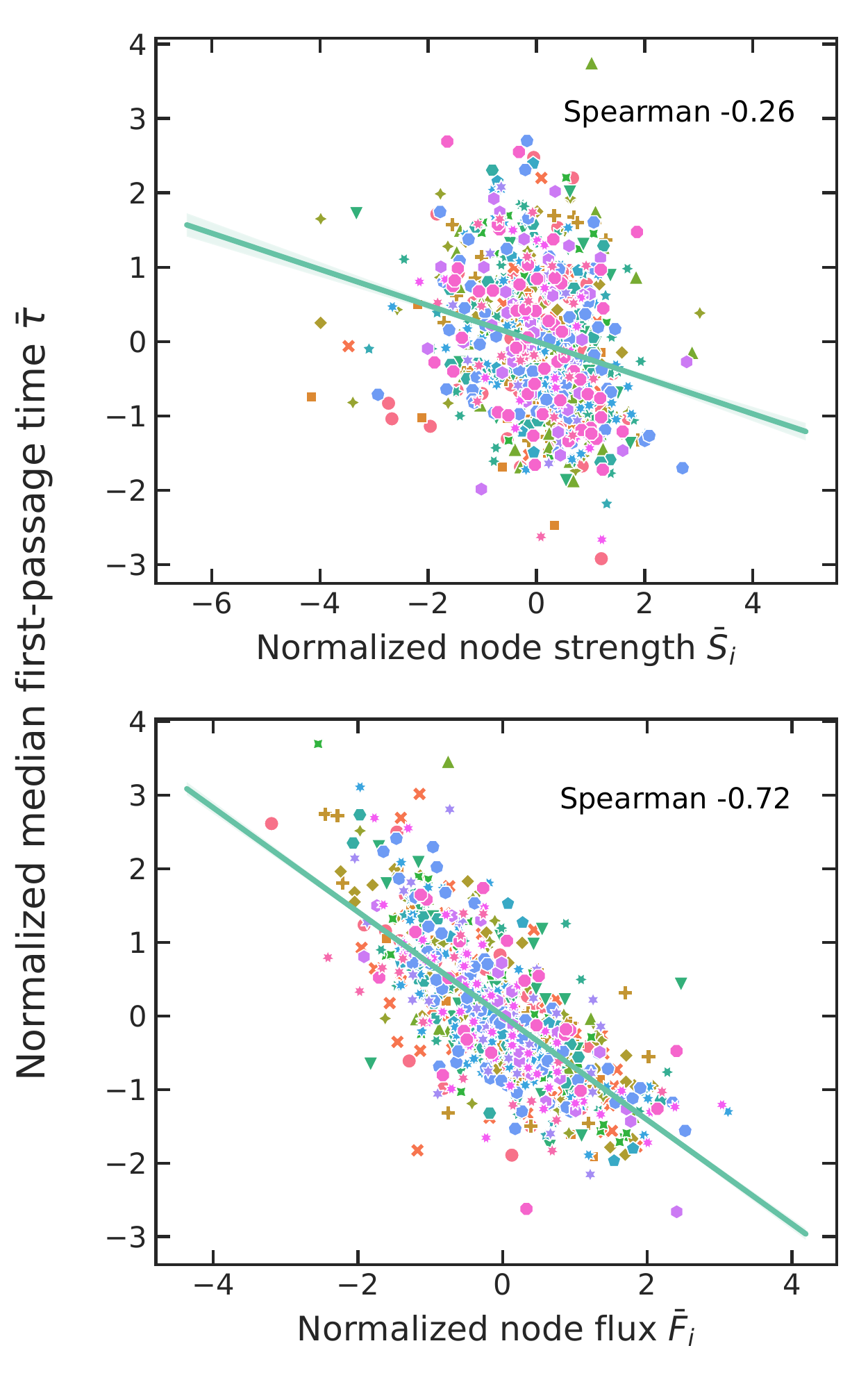} 
        \caption{%
        Median first-passage times $\bar\tau$ versus node strength $\bar {\cal S}_i$ (Eq.~\eqref{eq:node_strength}) (top) and node flux $\bar {\cal F}_i$  (Eq.~\eqref{eq:node_flux}) (bottom).  The bar-over-symbols ($\bar\tau$ and $\bar {\cal F}_i$) signpost normalized variables with respect to a Gaussian distribution (see App~\ref{app:rescaling_S_F_tau}). Each data point represents one consensus target bin, with colours and markers indicating different TFs. We show the top 20\% of TFs having the most variations in ${\cal F}_i$ and ${\cal S}_i$. The solid lines show the linear regressions and the corresponding 95\% CI (shaded area). 
        The data in both panels show negative correlations with $\bar \tau$ (see insets). However, the node flux has a stronger correlation than the node strength. By eye, we note that the goodness-of-fit varies between TFs, most showing a negative correlation.
        }%
        \label{fig:tau_vs_flux_str}
    \end{figure}
    
        \begin{figure*}
        \includegraphics[width=\textwidth]{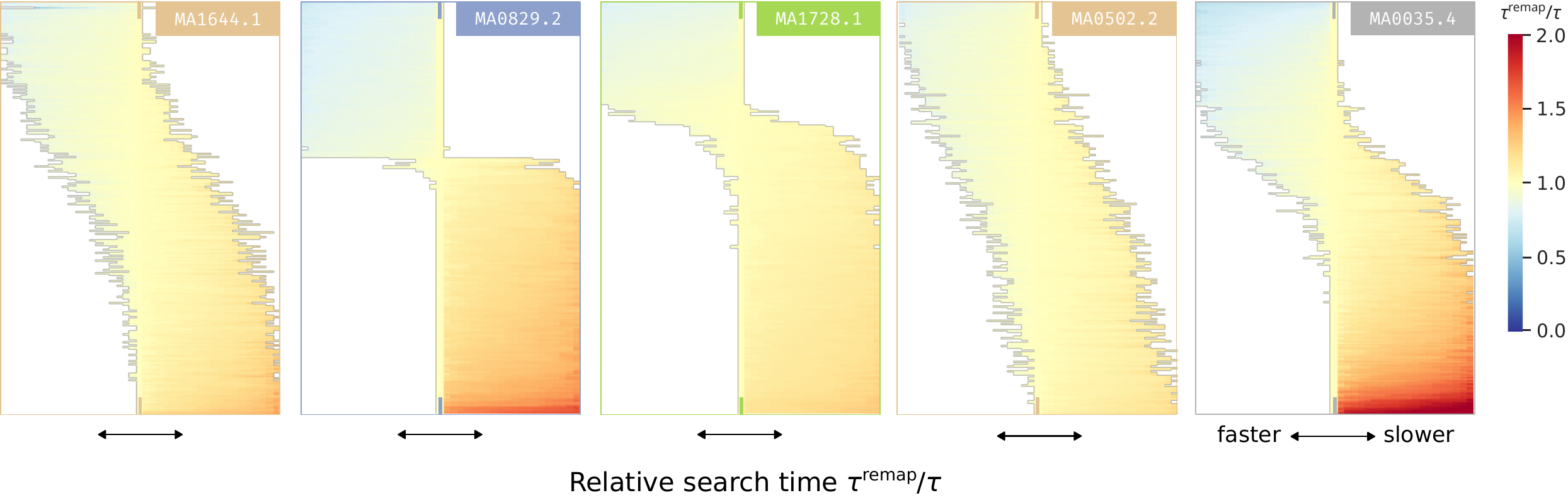} 
        \caption{%
        Median first-passage times for original and remapped systems associated with five TFs (one for each heat map). 
        Each heat map row represents one consensus target bin, where each pixel color, blue-to-red (see colour bar), indicate high to low median search time compared to the original configuration $\tau$.
        We sorted faster remaps ($\tau^{\rm remap} < \tau$) so that they fall to the left of the midpoint (note the tick marks) representing original (i.e., where $\tau^{\rm remap}/\tau=1$). Slower remaps $\tau^{\rm remap}> \tau$ end up on the right side. 
        For all TFs, we note a some consensus targets experience a substantial shift in $\tau^{\rm remap}$ when randomizing the sequences among the bins. 
        }%
        \label{fig:targetheatmap}
    \end{figure*}
    
    Last section showed that the median search times to individual consensus target bins vary significantly (Fig. \ref{fig:tau_sorted_6tfs}). To better understand why some target configurations have shorter search times than others, we analyzed the consensus-auxiliary binding site network. Particularly, we studied the link weights associated with nodes containing at least one consensus target. As we elaborated in Sec.~\ref{subsec:chromosome_wide_search}, we define the wight as the number of Hi-C contacts $v_{ij}$ [Eq.\eqref{eq:hicrates}].
   
    First, we analyzed the node strength, the sum of all link weights to node $i$, $\sum_j v_{ij}$. However, since the consensus bins act as absorbing targets, we exclude consensus-consensus connections and define the (partial) node strength as
    \begin{equation}\label{eq:node_strength}
        {\cal S}_i = \sum_{q} v_{iq}, \ \   q \neq {\rm consensus\, target\, bins}
    \end{equation}
    This metric reflects how tightly the rest of the network connects to consensus target bin $i$, and thereby help  directing searching proteins to designated chromatin regions (a large fraction of nodes contains at least one auxiliary binding site). 
    
    To better see how the node strengths changes with $\tau$, we renormalized the $ {\cal S}_i$ values associated with all consensus target bins and TFs and superimposed all data points in one plot [Fig.~\ref{fig:tau_vs_flux_str}(top)]. To renormalize the data, we transformed the node strengths and median first-passage times so that $\bar S_i$ and $\bar \tau$ belongs to a standard normal distribution [$N(0,1)$] after noticing that they are approximately distributed as a bell curve (App.~\ref{app:rescaling_S_F_tau}). In Fig.~\ref{fig:tau_vs_flux_str}(top), we show all data as a scatter plot, where each point represents a consensus target bin, and the symbols and colors indicate different TFs. Because the data forms a nearly vertical cloud, although with a minor negative tilt, we conclude that there is not a significant relationship between median-search times and node strength (the spearman correlation coefficient is -0.26).
    
    To find another network metric that better explains $\bar\tau$, we considered the node flux ${\cal F}_i$. This quantity considers the time spent in a bin before leaving in addition to the link weight. That is
    \begin{equation}\label{eq:node_flux}
       {\cal F}_i = \sum_q \frac{v_{iq}}{\tau_q} \ \   q \neq {\rm consensus\, target\, bins},
       \end{equation}
    where $\tau_q$ is median time until the TF escapes node $q$, that we denote persistence time. This means that each term $v_{iq}/\tau_q$ represents the effective rate of going from node $q$ to consensus node $i$ and the sum is the total rate.
    
    To estimate $\tau_q$, we simulated sequence-specific search until unbinding (by $k_{\rm off}$) in individual Hi-C bins, for all TFs (see App.~\ref{app:persistancetime}). Our simulations show that there is an exponential dependence between $\tau_q$ and the number of effective targets per bin, where the exact functional form is TF-dependent (see App.~\ref{app:effectivetargets}). So when calculating the node flux in Eq. \eqref{eq:node_flux}, we estimated $\tau_q$ from relationships like Fig. \ref{fig:persistencetime} and used $v_{iq}$ from the Hi-C data. 
      
    Next, we plotted the normalized node fluxes $\bar {\cal F}_i$ against the simulated median search time to consensus target bins $\bar{\tau}$. Compared to the node strengths $\bar {\cal S}_i$, we find a much stronger correlation: the Spearman coefficient is -0.72). Such a large correlation indicates that high node fluxes is a reliable predictor for fast search times. Like $\bar {\cal S}_i$ and $\bar \tau$, we normalized the node fluxes so that $\bar {\cal F}_i \sim N(0,1)$ (see App. \ref{app:rescaling_S_F_tau}).
    
\subsection{Optimal Network configurations for target-search}
    
    We found that the median search time depends on the node flux.
    Now we ask: could we achieve an even lower search time than in the original configuration if we rearranging the node flux weights? If so, by how much is the original ("wild type") deviating compared to other random counterparts?  
    
    While it is straightforward to scramble the escape rates in Eq.~\eqref{eq:node_flux} to calculate ${\cal F}_i$, we must do new simulations to get $\tau$ in systems with reshuffled persistence times. To his end, we randomized DNA sequences between the Hi-C bins while keeping the consensus node fixed. We call these randomized configurations "remapped", as we assign DNA sequences to Hi-C bins that differ from the standard genome sequence (e.g., hg 19 or hg 28). This is similar to TrIP--Chip, where experimenters insert DNA snippets at random DNA loci to study functional response to genomic positions \cite{kudo2010translational}.
    
    After randomizing the bin sequences, we simulated TF search, calculated median first-passage times, and compared these medians to the original arrangement. Using this remapping scheme, we found that the original configuration may differ as much as one order of magnitude relative to the randomized cases. Sometimes, the original configuration stands out as much slower or much faster compared to the remapped systems. We portray the data in Fig.~\ref{fig:targetheatmap}.
    
    Each figure panel show heatmaps of median search times to the consensus target bins for five TFs. Each row represents a sorted list of the median search times $\tau^{\rm remap}$ to one consensus target bin in 20 re-mapped configurations. The colors indicate the relative difference between the median first-passage times and the original configuration $\tau^{\rm remap}/\tau$. From the color spectrum, we note that the remaps vary as much as between different consensus target bins (Fig.~\ref{fig:tau_sorted_6tfs}) and even between different TFs (Fig.~\ref{fig:global_search}). The heatmaps show that the network positions of the consensus target bins are just as important as the TF or target type (e.g., sequence length).
    
    We also note that some remapped configurations are associated with much shorter or longer search time than the original configuration. This is best seen by following the heatmaps' midpoints that indicate where the ratio $\tau^{\rm remap}/\tau = 1$. All remaps associated with longer search times extend to the right of the midpoint and those having shorter are to the left. In most cases, the division is not half-way. Instead, the distribution is skewed, where the original may be fastest among the remaps (bottom row) and in other's it is the slowest (top rows). For example, in the second panel from the left (MA08292.2), the heatmap forms two blocks (one upper and one lower). The consensus bins in the upper block are hard to find when put in the original configuration. But after randomization, these move to more easy-to-find network locations, yielding $\tau^{\rm remap}<\tau$. The opposite happens in the lower block, where consensus bins become harder to find after randomization.
    
\begin{figure}
        \includegraphics[width=\columnwidth]{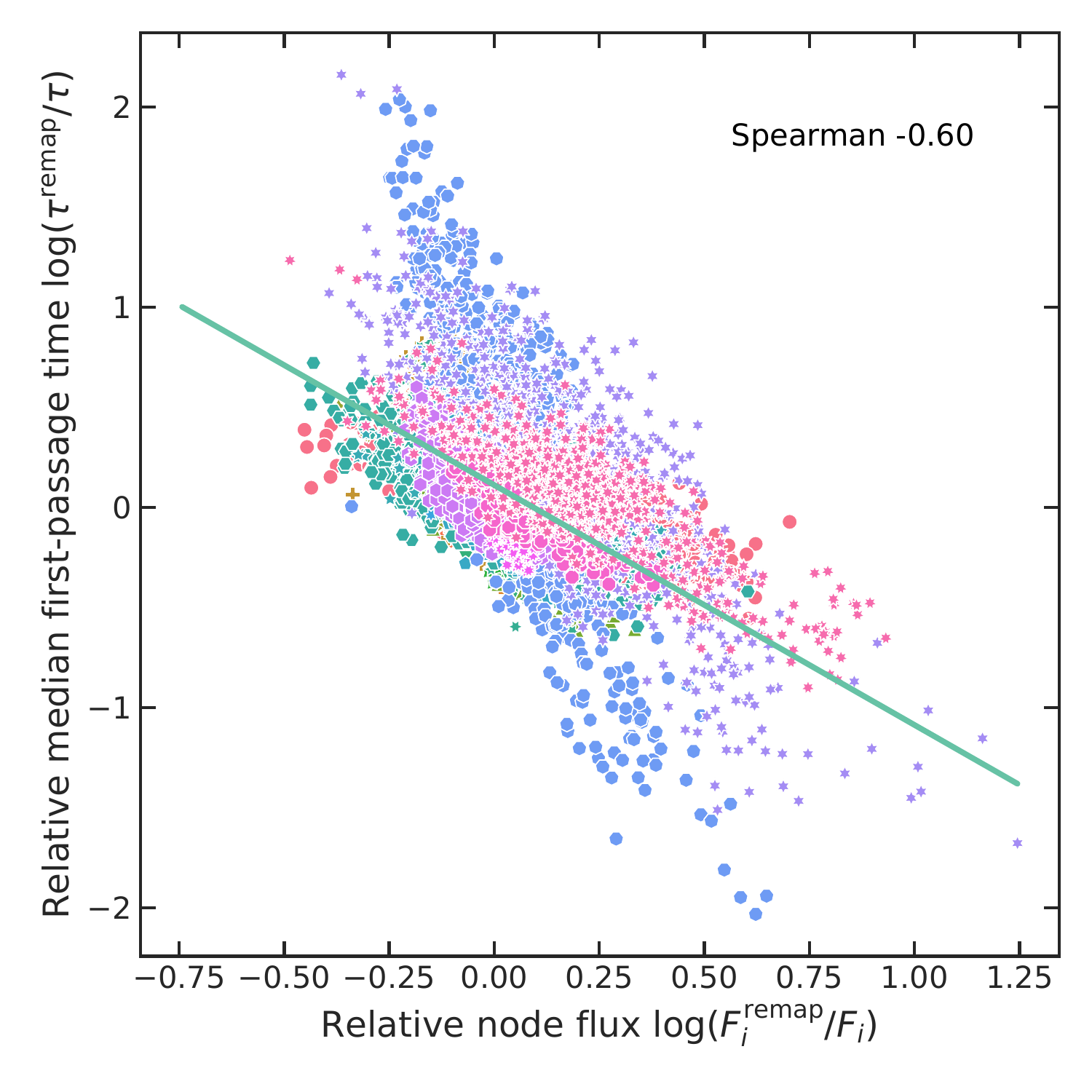} 
        \caption{%
        The relative change in search times $\tau^{\rm remap}/\tau$ and node flux ${\cal F}_i^{\rm remap}/{\cal F}_i$ for remapped systems. The plot shows the relative change in median first-passage times and node flux between the remapped systems and non-remapped system. Note the logged values. Each point represents a remap for a consensus target bin, i.e., 20 times as many points as in Fig.~\ref{fig:tau_vs_flux_str}, with colours and markers separating different TFs. The solid line shows a linear fit, with the corresponding 95\% CI (shaded area). As in Fig.~\ref{fig:tau_vs_flux_str}, we only look at the top 20\% of TFs.
        }%
        \label{fig:tau_vs_flux_remap}
    \end{figure}

    Finally, to validate that the relationship between median search time and node flux still holds (as we uncovered in Fig. \ref{fig:tau_vs_flux_str}), we computed ${\cal F}_i$ and $\tau$ for all remapped configurations ($\sim 10^4$ data points). We also re-scaled ${\cal F}_i$ and $\tau$ with the original configuration's values, $\tau^{\rm remap}/\tau$ and ${\cal F}_i^{\rm remap}/{\cal F}_i$. We plot these quantities in Fig.~\ref{fig:tau_vs_flux_remap}. As we expect, the data points (one TF per color) and the regression curve (green) show a negative correlation also for the remapped data, where the spearman correlation coefficient is slightly less negative than for the original configurations (-0.6 vs. -0.72). However, this correlation blends remapped data from multiple TFs, where some are more affected by remaps than others, note the blue dots compared to the other dots. Nevertheless, making separate regression lines, all show convincing negative slopes, albeit with slightly different tilts.
    
%% file: sections/conclusions.tex

    The most common mechanisms to regulate a gene is when a protein binds to a regulatory sequence to increase or decrease RNA transcription rates. However, these sequences are short relative to the genome length, so finding them poses a demanding search problem. Another problem is that there exist many nearly identical sequences. These almost-targets lie scattered across the genome, potentially causing proteins to get stuck. But as shown in a theoretical study in \textit{Escherichia coli}, such auxiliary binding sites may speed up the search if considering DNA looping. In this paper, we study if this phenomenon occurs on a chromosomes-wide scale, where DNA folding may help channel DNA-binding proteins, like transcription factors, to designated target regions. To this end, we develop a cross-scale approach that bridges established facilitated-diffusion models for the basepair-level search and chromosome-wide models capturing leaps spanning hundreds of millions of base pairs. As input to our cross-scale model, we used Hi-C data sets as a proxy for 3D proximity between long-ranged DNA segments (Hi-C bins) and binding profiles for more than 100 transcription factors. In most cases, transcription factors have $10-1000$ experimentally verified binding positions (consensus targets) and $\sim 10^3-10^4$ almost-targets (auxiliary binding sites) scattered across a chromosome. When coarse-grained into Hi-C bins, this amounts to $\sim 0-10$ targets or binding sites per bin (sometimes there are $>100$ auxiliary binding sites). 
    
    When studying the search to any Hi-C bin with at least one consensus target, we find that the targets' spatial distribution critically determines the variability among the transcription factors' median search time. But rather than occupying specific bins (nodes), it is more important how evenly the consensus targets spread over the DNA-contact network. Thus, considering the targets in aggregate, the network's connectivity and the spatial arrangement of the auxiliary binding sites seem to play a minor role in target search.
    
    This finding contrasts search times to single targets that likely are more relevant for gene-specific transcription regulation. For individual consensus-target bins, we found that the median search time is strongly associated with the node flux. The node flux generalizes the classical node strength (sum of weights to a node) as it combines the weight (i.e., the probability that two DNA segments form a long-ranged loop) and the dissociation rate (inverse persistence time). Across most transcription factors, high node flux is a strong indicator for short median search times. 
    
    We also found that the relative position of the bins containing auxiliary binding sites significantly affects search times for the consensus bins. To show this, we randomized the sequences among the Hi-C bins keeping the 3D contacts intact, and compared search times to the one for the actual target configuration. We noted a broad spectrum of outcomes. Sometimes, the natural arrangement was much faster than most random variations. In others, it was slower. It would be instructive to investigate the biological significance of fast versus slow configurations to better understand if the chromosomes' 3D organization funnels critical transcription factors to designated DNA regions. Since funneling seem to have evolved on the basepair level \cite{cencini2017}, similar driving forces could have shaped DNA's 3D organization. 
    
    Another aspect worthy of exploration is hidden targets. In some genomic regions, often epigenetically repressed, DNA (or chromatin) folds so densely that DNA-binding proteins cannot access and recognize the embedded sequence. Using DNAse data as a proxy for accessible DNA, one could use our cross-scale model to explore how omitting targets will affect median search times. We leave this study as an open problem for future work.
    
    To conclude, our work provides a framework for studying chromosome-wide search times for specific DNA sequences. Apart from transcription factors, we hope that our results and methods will help other researchers interested in DNA-protein binding associated with other genetic processes such as gene regulation, DNA repair \cite{esadze2018facilitated}, CRISPR/cas9 gene editing \cite{globyte2019crispr, lu2021search}, and epigenetics.

%% file: sections/appendix.tex
\section{Calculating the energy landscape}\label{app:energylandscape}
    
    To estimate the energy landscape that TFs experience when interacting with DNA, we employ standard methods (e.g., \cite{cencini2017}). These methods use the Position Frequency Matrix (PFM) containing the nucleotide occurrence at each position $i$ along the TFs target motif having length $L^{\mathrm{TF}}$. The PFM has four rows corresponding to each nucleotide type $s=\textrm{A, T, C, G}$, and $L^{\mathrm{TF}}$ columns each representing positions $i=0,\ldots ,L^{\mathrm{TF}}$.  We retrieved PFMs from the JASPAR database~\cite{jaspar}. 
    
    To get the TF binding profile $P^{\rm TF}(s, i)$, we converted the PFM  $n(s, i)$ into a TF-specific Position Probability Matrix (PPM). For some TFs there might be counts missing for one or more nucleotides at position $i$. This cause problems because $E\sim \log P$ gives infinite energies if $P=0$. To avoid this problem, we use additive smoothing~\cite{schutze2008introduction}, yeilding
    \begin{equation} \label{eq:pfm}
      P^{\rm TF}(s, i) = \frac{n^{\rm TF}(s, i) + p_s\pi(a)}
      {\sum_s (n^{\rm TF}(s, i) + p_s\pi(a))}, 
    \end{equation}
    where $p_s$ is the pseudocount (we use $p_s=1$) and $\pi(a) = 1/4$ is a uniform nucleotide probability distribution. 

    Next, we relate the PPM to the binding energy using Boltzmann's relationship~\cite{stormo1998specificity}
    \begin{equation}\label{eq:ppmtoe}
      \epsilon^{\rm TF}(s, i) = -\ln P^{\rm TF}(s, i).
    \end{equation} 
    where $k_B T = 1$. To calculate the interaction energy at position $x$, we sum over all positions $i$ in $L^{\rm TF}$ (see Fig.~\ref{fig:bp_schematic})
    \begin{equation} \label{eq:forward_complement}
      E^{\rm TF+}(x) = \sum_{i=0}^{L^{\rm TF} - 1} \epsilon^{\rm TF}(s, i+x),
    \end{equation} 

    While Eq.~\eqref{eq:forward_complement} calculates the forward looking energy (hence the superscript '+'), there is also reverse motif, $E^{\rm TF-}(x)$,  having a slightly different energy. Assuming that the TF associates randomly to either one side, we pick lowest binding energy, that is
    \begin{align}
        \begin{split}
              E^{\rm TF}(x) &= \min(E^{\rm TF+}(x), E^{\rm TF-}(x)) \\
                            &- \langle \min(E^{\rm TF+}(x), E^{\rm TF-}(x)) \rangle_x
        \end{split}
    \end{align}
    where the negation corresponds to setting the mean energy over the whole DNA sequence to zero \cite{cencini2017}.
    In our cross-scale model. we could include both the or forward and reverse energy landscapes separately. However, this would not lead to significant changes. 
    
\section{Normalizing the energy landscape}\label{app:genericsearcher}

\begin{figure}
        \includegraphics[width=\columnwidth]{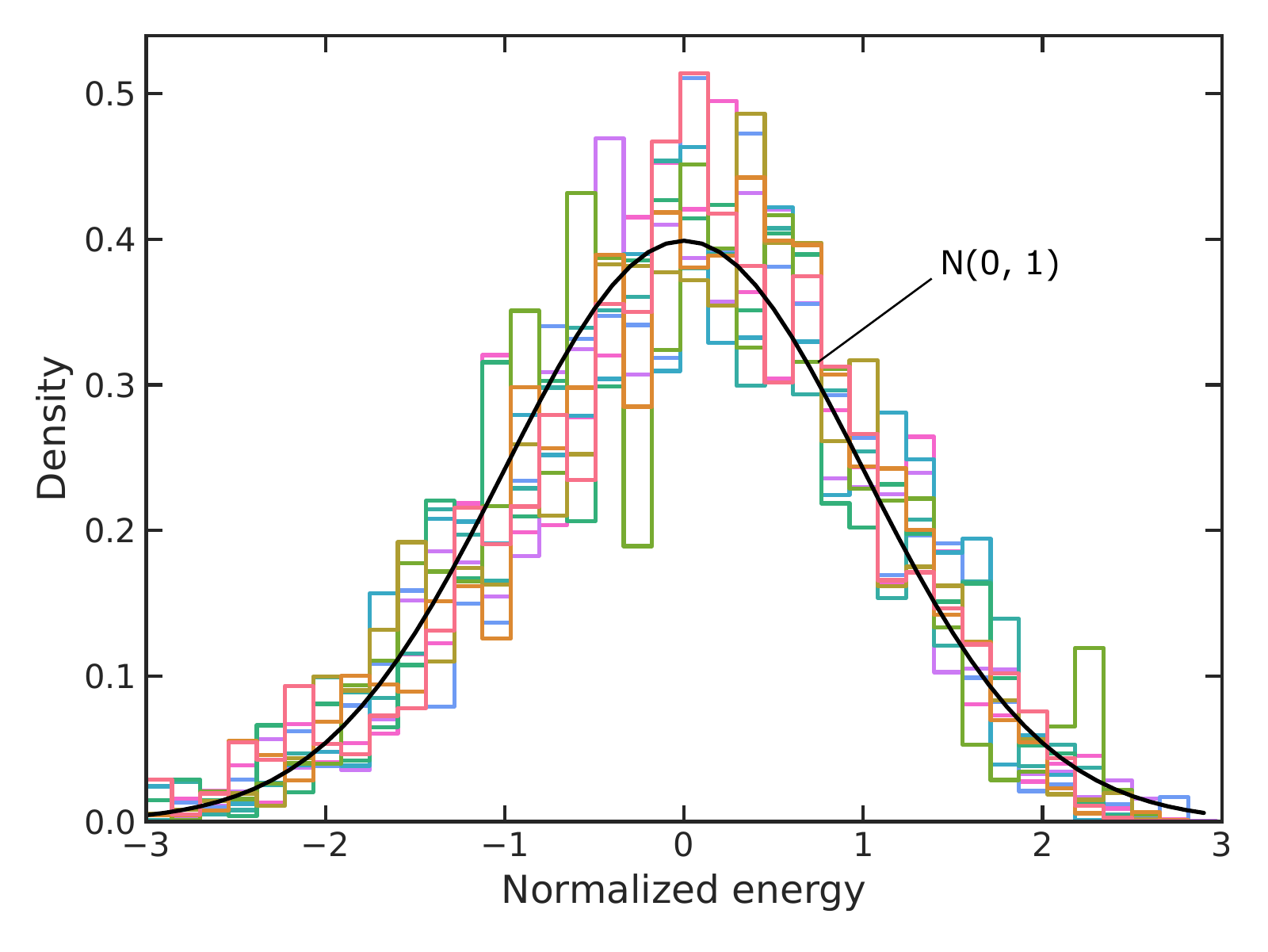}
        \caption{%
        Energy distributions of ten TFs normalized using Eq.~\eqref{eq:energyscaling} (one TF per color). The histogram includes $\sim 10^5$ values sampled randomly across the genome. The solid line shows the standard normal distribution, $N(0, 1)$.
        }%
        \label{fig:energynorm}
    \end{figure}

    Calculating the energy landscape as sum of varying variables (see App.~\ref{app:energylandscape}) results in a near Gaussian distribution of $E^{\rm TF}(x)$~\cite{mirny2009protein}. We therefore re-normalize the energy landscape according to
    \begin{equation} \label{eq:energyscaling}
        \bar{E}^{\rm TF}(x) = \frac{E^{\rm TF}(x) - \langle E^{\rm TF}(x)\rangle }{\sqrt{\langle {(E^{\rm TF}(x))}^2 \rangle - \langle E^{\rm TF}(x) \rangle^2}},
    \end{equation}
    so that $\bar{E}^{\rm TF}(x)\sim N(0, 1)$. We show energies for ten TFs in Fig~\ref{fig:energynorm}. 
    
    To calculate the corresponding free energy $\Delta G$ in the sequence-sensitive search model after normalization, we took the average of all $\Delta G$, i.e., $\Delta G = -3$.

\section{Median-first passage on dense networks}\label{app:tauonnormalizedhic}

    We showed in Sec.~\ref{subsec:global_search} that $\tau \sim \phi^{-1}$ for in a fully connected network. Here we generalize the derivation to dense but not fully connected networks (like Hi-C.)
    
    First, we denote the probability that the TF jumps to a target bin from from node $i$ as
    \begin{equation}
        P_i(1) = \sum_{j\in \Omega_t} \omega_{ij} = \phi_i,
    \end{equation}
    where $\Omega_t$ is the set of all target nodes. Formulated in this way, all $\phi_i$ are different but will fluctuate around some mean $\phi$, that is
    \begin{equation}
        \phi_i = \phi + \epsilon_i, \quad \epsilon_i \in [-\phi, 1-\phi].
    \end{equation}
    where $\epsilon_i$ is a random independent variable obeying
    \begin{equation}\label{eq:energy_correlations}
        \langle \epsilon_i \rangle = 0, \ \ \ \langle \epsilon_j \epsilon_i\rangle = \delta_{ij}
     \end{equation}
    where $\delta_{ij}$ is the kronecker delta and $ \langle . \rangle$ denotes ensemble average. Using these properties gives
    \begin{equation}
        \langle \phi_i \rangle = \langle \phi \rangle + \langle \epsilon_i \rangle = \phi.
    \end{equation}
    Second, we express the probability of making $n > 1$ jumps as
    \begin{equation}
        P_i(n) = \phi_1'\cdot \phi_2'\cdot \phi_3'\dots \phi_{n}
    \end{equation}
    Using $\phi' = (1-\phi)$ and $\phi_i' = \phi' - \epsilon_i$ yields
    \begin{widetext}
    \begin{align}
    \begin{split}
        P_i(n) 
                 &= (\phi' - \epsilon_1)\cdot(\phi' - \epsilon_2)\cdot(\phi' - \epsilon_3)\dots (\phi + \epsilon_{n}) \\
                 &= \left( (\phi')^{n-1} + (\phi')^{n-2}(-\epsilon_1 - \epsilon_2 - \dots) + (\phi')^{n-3}(\epsilon_1\epsilon_2 + \dots) + \dots\right)\cdot(\phi + \epsilon_{n}).
    \end{split}
    \end{align}
    \end{widetext}
    Averaging over all positions $i$ and using Eq. \eqref{eq:energy_correlations} gives
    \begin{align}
    \begin{split}
        \langle P_i(n) \rangle &= \langle (\phi')^{n-1} \dots \rangle \langle \phi + \epsilon_{n} \rangle \\
                               &= \phi'^{n-1} \phi = (1-\phi)^{n-1} \phi,
    \end{split}
    \end{align}
    This equation is similar to Eq. \eqref{eq:P_i} from which the rest of the argumentation leading up to $\tau \sim \phi^{-1}$ is identical.
    
\section{Effective binding sites}\label{app:effectivetargets}
    
    \begin{figure}[htpb]
        \centering
        \includegraphics[width=0.9\linewidth]{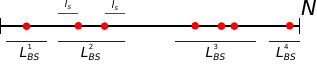}
        \caption{%
        Schematic of how close-by binding site reduce their effectiveness. Here we show a list of $N$ sites with auxiliary binding site marked with red dots. For each cluster of binding site, the window $L^i_{\rm BS}$ stretches $l_s$ from the outermost binding site. If a TF lands within this region, it will most probably hit one of the binding sites. However, if $L^i_{\rm BS} \approx l_s$, then the cluster effectively only act as one binding site.
        }%
        \label{fig:effectivetargets}
    \end{figure}
    Auxiliary binding sites sequester the TF stronger than other positions along the sequence. If close together, such sites my act as one if they lie within the TFs sliding length $l_s$ (see Fig.~\ref{fig:effectivetargets}). In this section, we calculate the relationship between the number of effective binding sites ($\Sigma_{\rm eff}$) and the number of actual binding sites ($\Sigma$) across the a Hi-C bins. As in Fig.~\ref{fig:effectivetargets} we define regions $L_{\rm BS}^i$ of length $2l_s$ centered around each binding site. To calculate the number of effective binding sites, we sum $L_{\rm BS}^i$ and divide by twice the sliding length
    \begin{equation} \label{eq:effectivetargets}
        \Sigma_{\rm eff} = \left\lfloor\frac{1}{2l_s}\sum_i L_{\rm BS}^i + \frac{1}{2}\right\rfloor,
    \end{equation}
    where $\lfloor \cdot + 1/2 \rfloor$ denotes rounding half up ($\lfloor \cdot \rfloor$ is the round-down function). As we merge $L_{\rm BS}^i$, the limit is easy to compute
    \begin{equation}
        \lim_{\Sigma\rightarrow\infty} \Sigma_{\rm eff} 
        = \lim_{\Sigma\rightarrow\infty}\left\lfloor\frac{1}{2l_s}\sum_i L_{\rm BS}^i +
        \frac{1}{2}\right\rfloor
        = \frac{N}{2l_s}.
    \end{equation}
    
    We calculate this measure analytically by splitting the region of length $N$ into equally sized boxes $N_b=N/2l_s$. Next, if randomly placing $\Sigma$ binding sites, the expected number of boxes with only one binding sites is
    \begin{equation} \label{eq:effectivetargetstheory}
        \Sigma_{\rm eff}(\Sigma) = \sum_{i=0}^{\Sigma - 1} {(-1)}^i \binom{\Sigma}{i+1} \frac{1}{{(N_b)}^{i}}.
    \end{equation}
    If the size of each box is $N_b = 2l_s$, two binding sites randomly placed in the same box corresponds to two binding sites sharing one $L_{\rm BS}$. As such, Eq.~\eqref{eq:effectivetargetstheory} approximately corresponds to the number of effective binding sites as given by Eq.~\eqref{eq:effectivetargets}. We plot Eq.~\eqref{eq:effectivetargetstheory} for two different sliding lengths against the actual results from data. We can see that the real data is more clustered than expected, yielding a `effective' sliding length that is longer than the actual sliding length.
    \begin{figure}[htpb]
        \centering
        \includegraphics[width=\columnwidth]{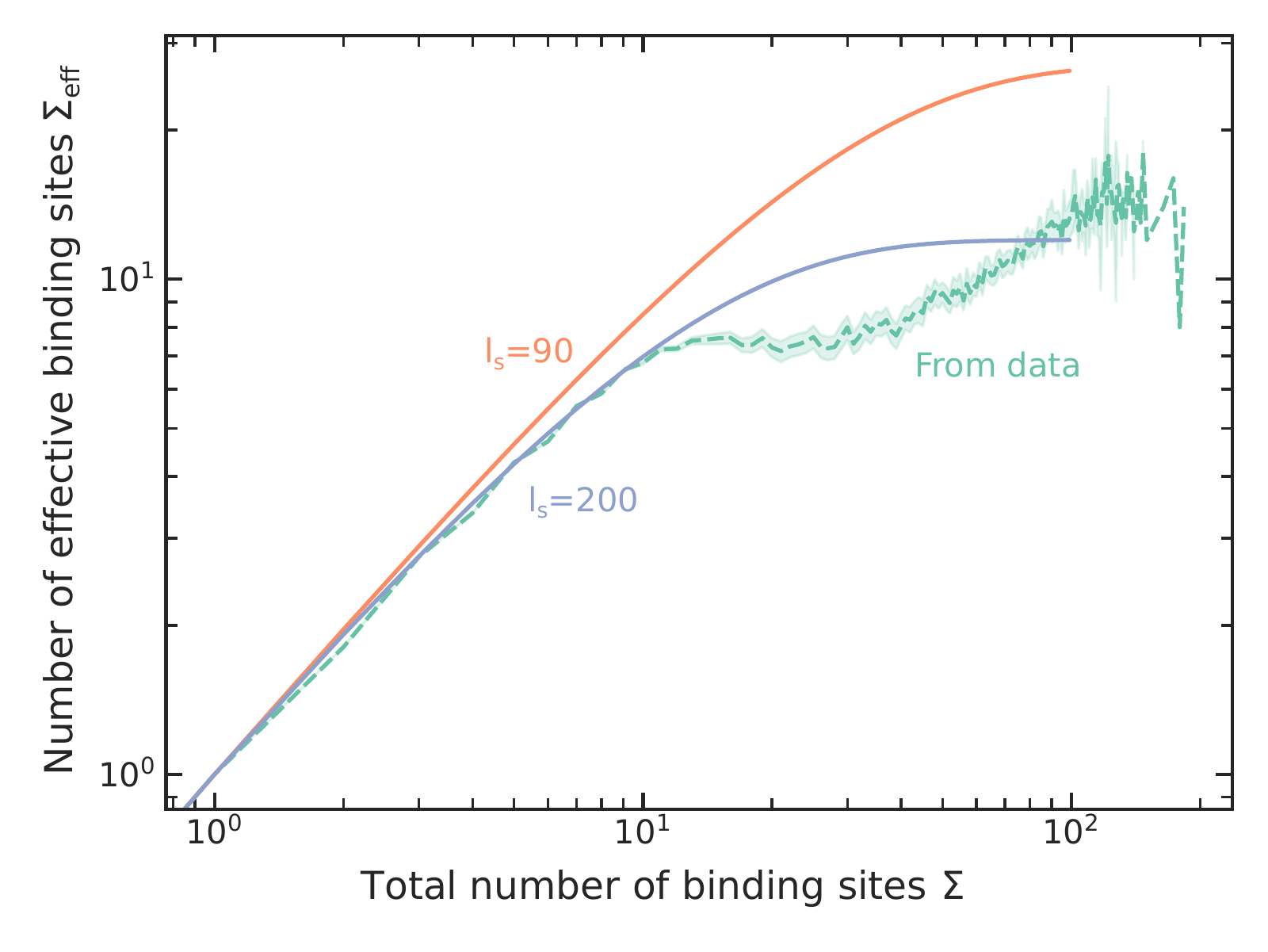}
        \caption{%
        The number of effective binding sites based on the number of binding sites, comparing data and theory. The two lines in orange and blue are calculated using Eq.~\eqref{eq:effectivetargetstheory} using two different sliding lengths $l_s$. The green line is calculated from binding site placement identified using MOODS for $~100$ TFs, see Sec.~\ref{subsubsec:moods}. We calculated the number of effective binding sites using the mean sliding length of $90$bp. The curve from data does not follow the $90$bp curve, but rather the $200$bp curve, indicating that binding sites on the chromosome is more clustered than just by random placement \cite{cencini2017}.
        }%
        \label{fig:persistencetime}
    \end{figure}
    
\section{Residence time on 5 kb regions}\label{app:persistancetime}
    
    In App. \ref{app:effectivetargets}, we quantified the number of effective binding sites in each Hi-C bin. Here we quantify the relationship between the number of effective binding sites and the persistence time $t_q$, i.e., the time TFs spend in a Hi-C bin before disassociating. 
    
\begin{figure}
        \centering
        \includegraphics[width=\columnwidth]{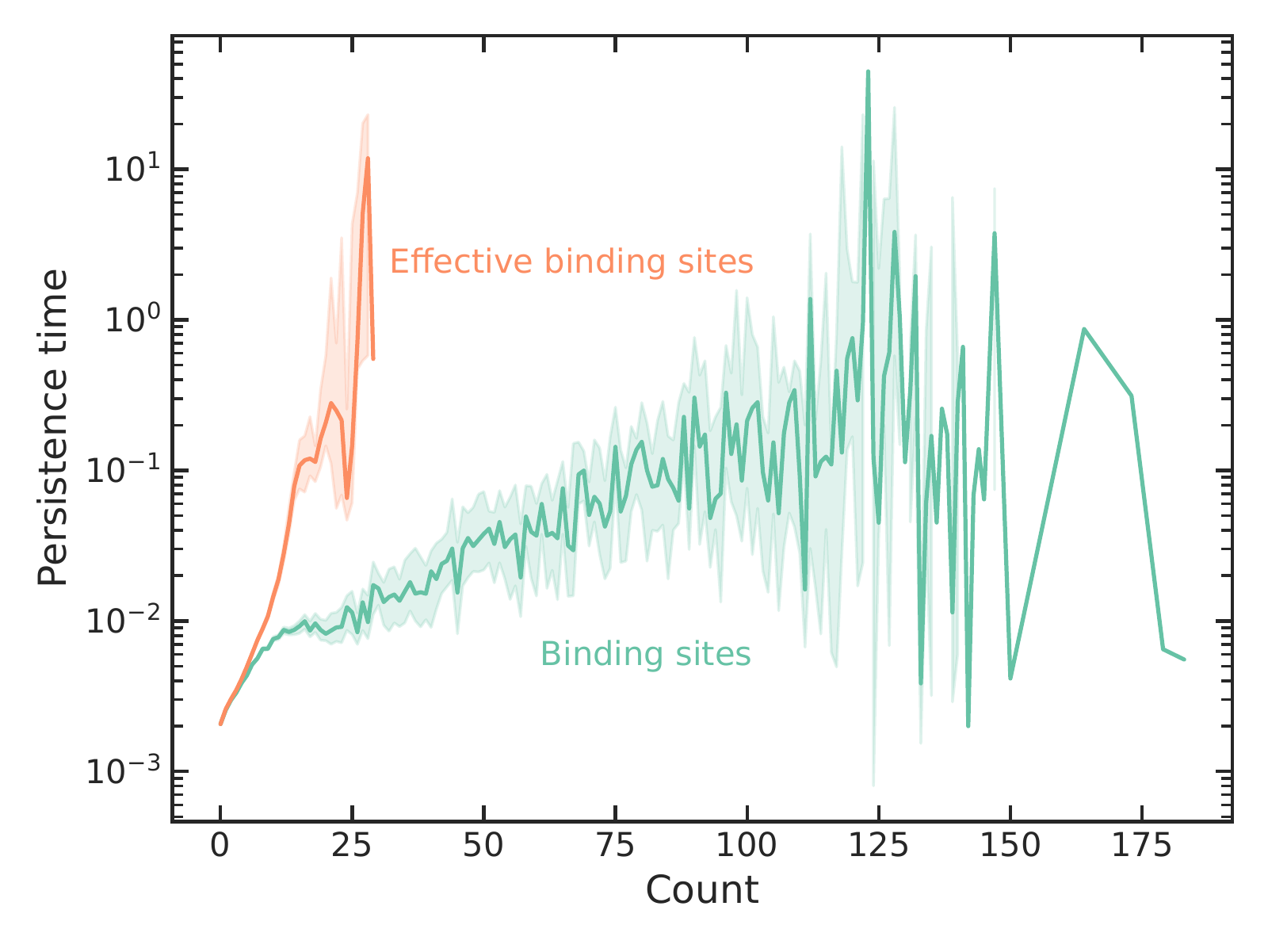}
        \caption{%
        The persistence time $t_q$ for Hi-C bins versus two binding site counts (total and effective number of binding sites). The persistence time is calculated from $~100$ simulations per Hi-C bin. Each point is aggregated from all bins and all TFs. We see that the green line, which just counts the number of binding sites in each bin, shows a weak increase of the persistence time as the count increases. However, in the orange line, which considers effective binding sites, we see a much stronger increase as the binding site count goes up.
        }%
        \label{fig:nefftdt}
    \end{figure}

    Using our simulation framework, we place a TF in a random Hi-C bin and simulate sequence-specific search until unbinding by $k_{\rm off}$ (see Sec.~\ref{subsec:sequence_sensitive}). In Fig.~\ref{fig:nefftdt}, we plot $t_q$ versus two measures of binding site count, noting that both metrics show an apparent increase in persistence time as the count increases. As explained in App.~\ref{app:effectivetargets}, only counting the number of binding sites gives a weaker increase. However, looking at the number of effective binding sites, we see a more substantial persistence time growth. This shows that clustered binding sites are less effective in sequestering TFs than when spread out.
    
\section{Distribution of node flux, node strength, and the median persistence time}
\label{app:rescaling_S_F_tau}
    
\begin{figure*} \label{fig:renormalization}
        \centering
        \includegraphics[width=\textwidth]{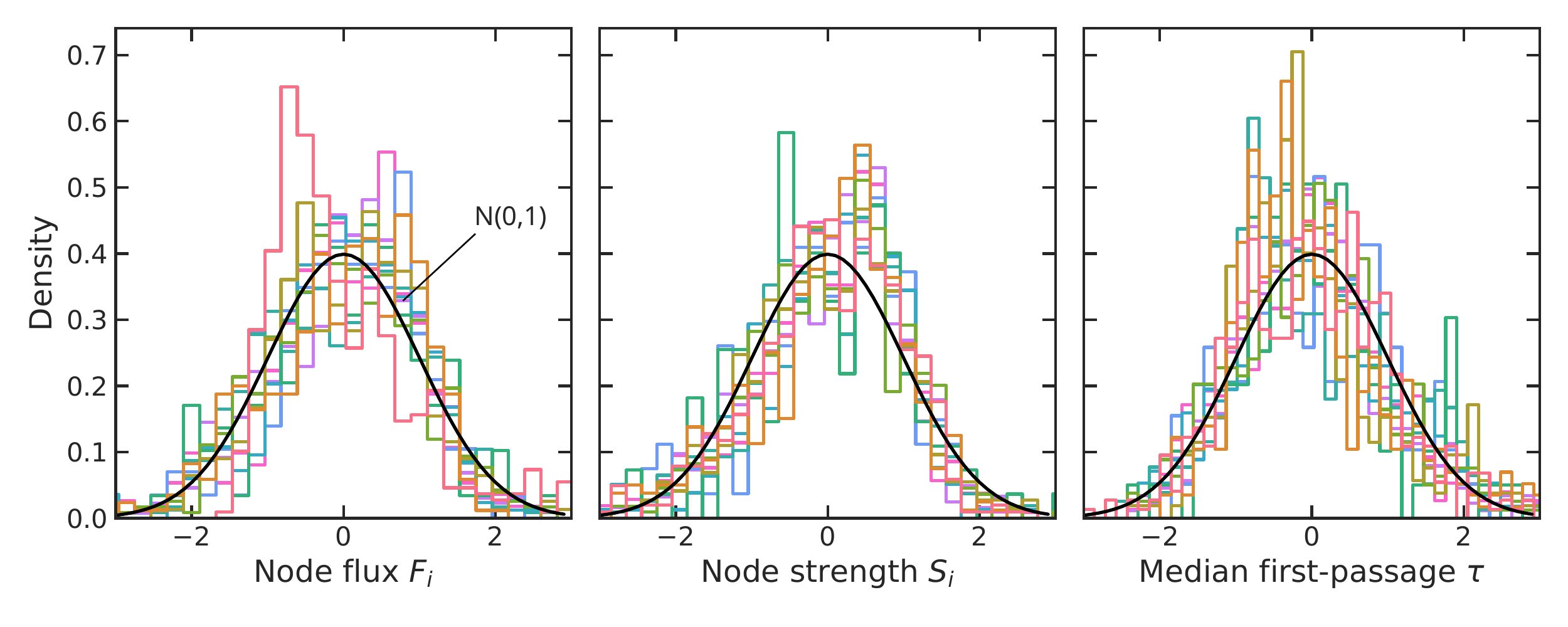}
        \caption{%
        Density distributions of median first-passage time $\tau$, node strength ${\cal S}_i$ and node flux ${\cal F}_i$ after gaussian renormalization [Eq. \eqref{eq:renormalization}]. Each colour represents a different TF. 
        As solid line, we show the standard normal distribution $N(0, 1)$.
        }%
        \label{fig:hists_norm_tau_str_flux}
    \end{figure*}

    To simplify the visualization among many TFs, we renormalized $S_i$, $F_i$ and $\tau$. As we show in Fig. \ref{fig:renormalization}, all these varibale collabse to a standard normal distribution after the transformation
    \begin{equation}\label{eq:renormalization}
        (X-\langle X\rangle)/\langle X^2 \rangle + \langle X \rangle^2, \ \ X= \tau, {\cal S}_i, {\cal F}_i
    \end{equation}
    
%% file: manuscript.bbl
\begin{thebibliography}{10}

\bibitem{Note1}
Typical Transcription factor sequences are 10-50 bp long. The human genome has
  $7\cdot 10^9$ bps.

\bibitem{hammar2012lac}
Petter Hammar, Prune Leroy, Anel Mahmutovic, Erik~G Marklund, Otto~G Berg, and
  Johan Elf.
\newblock The lac repressor displays facilitated diffusion in living cells.
\newblock {\em Science}, 336(6088):1595--1598, 2012.

\bibitem{marklund2022sequence}
Emil Marklund, Guanzhong Mao, Jinwen Yuan, Spartak Zikrin, Eldar Abdurakhmanov,
  Sebastian Deindl, and Johan Elf.
\newblock Sequence specificity in dna binding is mainly governed by
  association.
\newblock {\em Science}, 375(6579):442--445, 2022.

\bibitem{bauer2015real}
Maximilian Bauer, Emil~S Rasmussen, Michael~A Lomholt, and Ralf Metzler.
\newblock Real sequence effects on the search dynamics of transcription factors
  on dna.
\newblock {\em Scientific Reports}, 5:10072, 2015.

\bibitem{cortini2018theoretical}
Ruggero Cortini and Guillaume~J Filion.
\newblock Theoretical principles of transcription factor traffic on folded
  chromatin.
\newblock {\em Nature communications}, 9(1):1--10, 2018.

\bibitem{kvon2012hot}
Evgeny~Z Kvon, Gerald Stampfel, J~Omar Y{\'a}{\~n}ez-Cuna, Barry~J Dickson, and
  Alexander Stark.
\newblock Hot regions function as patterned developmental enhancers and have a
  distinct cis-regulatory signature.
\newblock {\em Genes \& development}, 26(9):908--913, 2012.

\bibitem{brackley2012facilitated}
Chris~A Brackley, Mike~E Cates, and Davide Marenduzzo.
\newblock Facilitated diffusion on mobile dna: configurational traps and
  sequence heterogeneity.
\newblock {\em Physical review letters}, 109(16):168103, 2012.

\bibitem{rao20143d}
Suhas~SP Rao, Miriam~H Huntley, Neva~C Durand, Elena~K Stamenova, Ivan~D
  Bochkov, James~T Robinson, Adrian~L Sanborn, Ido Machol, Arina~D Omer, Eric~S
  Lander, et~al.
\newblock A 3d map of the human genome at kilobase resolution reveals
  principles of chromatin looping.
\newblock {\em Cell}, 159(7):1665--1680, 2014.

\bibitem{fornes2020jaspar}
Oriol Fornes, Jaime~A Castro-Mondragon, Aziz Khan, Robin Van~der Lee, Xi~Zhang,
  Phillip~A Richmond, Bhavi~P Modi, Solenne Correard, Marius Gheorghe, Damir
  Barana{\v{s}}i{\'c}, et~al.
\newblock Jaspar 2020: update of the open-access database of transcription
  factor binding profiles.
\newblock {\em Nucleic acids research}, 48(D1):D87--D92, 2020.

\bibitem{korhonen2009moods}
Janne Korhonen, Petri Martinm{\"a}ki, Cinzia Pizzi, Pasi Rastas, and Esko
  Ukkonen.
\newblock Moods: fast search for position weight matrix matches in dna
  sequences.
\newblock {\em Bioinformatics}, 25(23):3181--3182, 2009.

\bibitem{cencini2017}
Massimo Cencini and Simone Pigolotti.
\newblock Energetic funnel facilitates facilitated diffusion.
\newblock {\em Nucleic Acids Research}, 46(3):558--567, 2017.

\bibitem{benichou2011intermittent}
Olivier B{\'e}nichou, Claude Loverdo, Michel Moreau, and Raphael Voituriez.
\newblock Intermittent search strategies.
\newblock {\em Reviews of Modern Physics}, 83(1):81, 2011.

\bibitem{lomholt2009facilitated}
Michael~A Lomholt, Bram van~den Broek, Svenja-Marei~J Kalisch, Gijs~JL Wuite,
  and Ralf Metzler.
\newblock Facilitated diffusion with dna coiling.
\newblock {\em Proceedings of the National Academy of Sciences},
  106(20):8204--8208, 2009.

\bibitem{mirny2009protein}
Leonid Mirny, Michael Slutsky, Zeba Wunderlich, Anahita Tafvizi, Jason Leith,
  and Andrej Kosmrlj.
\newblock How a protein searches for its site on dna: the mechanism of
  facilitated diffusion.
\newblock {\em Journal of Physics A: Mathematical and Theoretical},
  42(43):434013, 2009.

\bibitem{felipe2021dna}
Cayke Felipe, Jaeoh Shin, and Anatoly~B Kolomeisky.
\newblock Dna looping and dna conformational fluctuations can accelerate
  protein target search.
\newblock {\em The Journal of Physical Chemistry B}, 125(7):1727--1734, 2021.

\bibitem{nyberg2021modeling}
Markus Nyberg, Tobias Ambj{\"o}rnsson, Per Stenberg, and Ludvig Lizana.
\newblock Modeling protein target search in human chromosomes.
\newblock {\em Physical Review Research}, 3(1):013055, 2021.

\bibitem{bagchi2008diffusion}
Biman Bagchi, Paul~C Blainey, and X~Sunney Xie.
\newblock Diffusion constant of a nonspecifically bound protein undergoing
  curvilinear motion along dna.
\newblock {\em The Journal of Physical Chemistry B}, 112(19):6282--6284, 2008.

\bibitem{murugan2010theory}
R~Murugan.
\newblock Theory of site-specific dna-protein interactions in the presence of
  conformational fluctuations of dna binding domains.
\newblock {\em Biophysical Journal}, 99(2):353--359, 2010.

\bibitem{Note2}
If increasing $\rho $ beyond 0.3, the diffusion constant becomes too low than
  observed in experiments \cite {cencini2017}).

\bibitem{winter1981diffusion}
Robert~B Winter, Otto~G Berg, and Peter~H Von~Hippel.
\newblock Diffusion-driven mechanisms of protein translocation on nucleic
  acids. 3. the escherichia coli lac repressor-operator interaction: kinetic
  measurements and conclusions.
\newblock {\em Biochemistry}, 20(24):6961--6977, 1981.

\bibitem{dixon2012topological}
Jesse~R Dixon, Siddarth Selvaraj, Feng Yue, Audrey Kim, Yan Li, Yin Shen, Ming
  Hu, Jun~S Liu, and Bing Ren.
\newblock Topological domains in mammalian genomes identified by analysis of
  chromatin interactions.
\newblock {\em Nature}, 485(7398):376--380, 2012.

\bibitem{lee2019mapping}
Sang~Hoon Lee, Yeonghoon Kim, Sungmin Lee, Xavier Durang, Per Stenberg,
  Jae-Hyung Jeon, and Ludvig Lizana.
\newblock Mapping the spectrum of 3d communities in human chromosome
  conformation capture data.
\newblock {\em Scientific reports}, 9(1):1--7, 2019.

\bibitem{Note3}
We acknowledge that the genome-averaged contact probability in human decays as
  $l^{-1.08}$ \cite {lieberman2009comprehensive} or $l^{-0.75}$ within TADs
  \cite {sanborn2015chromatin}. However, individual pairwise contacts show
  significant deviations from these average relationships.

\bibitem{farnham2009insights}
Peggy~J Farnham.
\newblock Insights from genomic profiling of transcription factors.
\newblock {\em Nature Reviews Genetics}, 10(9):605--616, 2009.

\bibitem{jaspar}
Jaspar 2020: An open-access database of transcription factor binding profiles.

\bibitem{jaspardocs}
Jaspar documentation.
\newblock \url{https://jaspar.genereg.net/docs/}.
\newblock Accessed: 2022-11-07.

\bibitem{stormo1982use}
Gary~D Stormo, Thomas~D Schneider, Larry Gold, and Andrzej Ehrenfeucht.
\newblock Use of the ‘perceptron’algorithm to distinguish translational
  initiation sites in e. coli.
\newblock {\em Nucleic acids research}, 10(9):2997--3011, 1982.

\bibitem{slutsky2004kinetics}
Michael Slutsky and Leonid~A Mirny.
\newblock Kinetics of protein-dna interaction: facilitated target location in
  sequence-dependent potential.
\newblock {\em Biophysical journal}, 87(6):4021--4035, 2004.

\bibitem{van2008dna}
Bram van~den Broek, Michael~Andersen Lomholt, S-MJ Kalisch, Ralf Metzler, and
  Gijs~JL Wuite.
\newblock How dna coiling enhances target localization by proteins.
\newblock {\em Proceedings of the National Academy of Sciences},
  105(41):15738--15742, 2008.

\bibitem{amitai2018chromatin}
Assaf Amitai.
\newblock Chromatin configuration affects the dynamics and distribution of a
  transiently interacting protein.
\newblock {\em Biophysical journal}, 114(4):766--771, 2018.

\bibitem{hu2006proteins}
Tao Hu, A~Yu Grosberg, and BI~Shklovskii.
\newblock How proteins search for their specific sites on dna: the role of dna
  conformation.
\newblock {\em Biophysical journal}, 90(8):2731--2744, 2006.

\bibitem{lomholt2005optimal}
Michael~A Lomholt, Tobias Ambj{\"o}rnsson, and Ralf Metzler.
\newblock Optimal target search on a fast-folding polymer chain with volume
  exchange.
\newblock {\em Physical review letters}, 95(26):260603, 2005.

\bibitem{smrek2015facilitated}
Jan Smrek and Alexander~Y Grosberg.
\newblock Facilitated diffusion of proteins through crumpled fractal dna
  globules.
\newblock {\em Physical Review E}, 92(1):012702, 2015.

\bibitem{kent2002human}
W~James Kent, Charles~W Sugnet, Terrence~S Furey, Krishna~M Roskin, Tom~H
  Pringle, Alan~M Zahler, and David Haussler.
\newblock The human genome browser at ucsc.
\newblock {\em Genome research}, 12(6):996--1006, 2002.

\bibitem{edgar2002gene}
Ron Edgar, Michael Domrachev, and Alex~E Lash.
\newblock Gene expression omnibus: Ncbi gene expression and hybridization array
  data repository.
\newblock {\em Nucleic acids research}, 30(1):207--210, 2002.

\bibitem{knight2013fast}
Philip~A Knight and Daniel Ruiz.
\newblock A fast algorithm for matrix balancing.
\newblock {\em IMA Journal of Numerical Analysis}, 33(3):1029--1047, 2013.

\bibitem{kumar2017genome}
Rajendra Kumar, Haitham Sobhy, Per Stenberg, and Ludvig Lizana.
\newblock Genome contact map explorer: a platform for the comparison,
  interactive visualization and analysis of genome contact maps.
\newblock {\em Nucleic acids research}, 45(17):e152--e152, 2017.

\bibitem{Note4}
Close to 80\% of the genes are active, which is significantly more that the
  other chromosomes where about 50\% are active \cite {nyberg2021modeling}.

\bibitem{kaufmann2015inter}
Stefanie Kaufmann, Christiane Fuchs, Mariya Gonik, Ekaterina~E Khrameeva,
  Andrey~A Mironov, and Dmitrij Frishman.
\newblock Inter-chromosomal contact networks provide insights into mammalian
  chromatin organization.
\newblock {\em PloS one}, 10(5):e0126125, 2015.

\bibitem{letter-value-plot}
Heike Hofmann, Karen Kafadar, and Hadley Wickham.
\newblock Letter-value plots: Boxplots for large data.
\newblock Technical report, had.co.nz, 2011.

\bibitem{gillespie1976general}
Daniel~T Gillespie.
\newblock A general method for numerically simulating the stochastic time
  evolution of coupled chemical reactions.
\newblock {\em Journal of computational physics}, 22(4):403--434, 1976.

\bibitem{Note5}
We implemented intermittent power-law jumps within the 5kb region too, but we
  did not notice any meaningful differences when studying large-distance search
  times.

\bibitem{Note6}
A full derivation not assuming a fully-connected network is shown in App~\ref
  {app:tauonnormalizedhic}.

\bibitem{raade1999mathematics}
Lennart R{\aa}de and Bertil Westergren.
\newblock {\em Mathematics handbook for science and engineering}, volume~5.
\newblock Springer, 1999.

\bibitem{kudo2010translational}
Kenji Kudo, Yaguang Xi, Yuan Wang, Bo~Song, Edward Chu, Jingyue Ju, James~J
  Russo, and Jingfang Ju.
\newblock Translational control analysis by translationally active rna
  capture/microarray analysis (trip--chip).
\newblock {\em Nucleic acids research}, 38(9):e104--e104, 2010.

\bibitem{esadze2018facilitated}
Alexandre Esadze and James~T Stivers.
\newblock Facilitated diffusion mechanisms in dna base excision repair and
  transcriptional activation.
\newblock {\em Chemical reviews}, 118(23):11298--11323, 2018.

\bibitem{globyte2019crispr}
Viktorija Globyte, Seung~Hwan Lee, Taegeun Bae, Jin-Soo Kim, and Chirlmin Joo.
\newblock Crispr/cas9 searches for a protospacer adjacent motif by lateral
  diffusion.
\newblock {\em The EMBO journal}, 38(4):e99466, 2019.

\bibitem{lu2021search}
Qiao Lu, Deepak Bhat, Darya Stepanenko, Simone Pigolotti, et~al.
\newblock Search and localization dynamics of the crispr-cas9 system.
\newblock {\em Physical Review Letters}, 127(20):208102, 2021.

\bibitem{schutze2008introduction}
Hinrich Sch{\"u}tze, Christopher~D Manning, and Prabhakar Raghavan.
\newblock {\em Introduction to information retrieval}, volume~39.
\newblock Cambridge University Press Cambridge, 2008.

\bibitem{stormo1998specificity}
Gary~D Stormo and Dana~S Fields.
\newblock Specificity, free energy and information content in protein--dna
  interactions.
\newblock {\em Trends in biochemical sciences}, 23(3):109--113, 1998.

\bibitem{lieberman2009comprehensive}
Erez Lieberman-Aiden, Nynke~L Van~Berkum, Louise Williams, Maxim Imakaev,
  Tobias Ragoczy, Agnes Telling, Ido Amit, Bryan~R Lajoie, Peter~J Sabo,
  Michael~O Dorschner, et~al.
\newblock Comprehensive mapping of long-range interactions reveals folding
  principles of the human genome.
\newblock {\em science}, 326(5950):289--293, 2009.

\bibitem{sanborn2015chromatin}
Adrian~L Sanborn, Suhas~SP Rao, Su-Chen Huang, Neva~C Durand, Miriam~H Huntley,
  Andrew~I Jewett, Ivan~D Bochkov, Dharmaraj Chinnappan, Ashok Cutkosky, Jian
  Li, et~al.
\newblock Chromatin extrusion explains key features of loop and domain
  formation in wild-type and engineered genomes.
\newblock {\em Proceedings of the National Academy of Sciences},
  112(47):E6456--E6465, 2015.

\end{thebibliography}
